\documentclass[11pt]{article}
\usepackage{graphicx}
\input{epsf.sty}
\usepackage{amsmath}

\begin{document}

\hfill{UTTG-17-06}

\vspace{36pt}

\begin{center}
{\large {\bf {Quantum Cosmological Correlations in an Inflating
Universe: Can fermion and gauge fields loops give a scale free
spectrum?}}}

\vspace{36pt} Kanokkuan Chaicherdsakul\footnote{Electronic
address:
kanokkua@physics.utexas.edu}\\
{\em Theory Group, Department of Physics, University of
Texas\\
Austin, TX, 78712}

\vspace{30pt}

\noindent
{\bf Abstract}
\end{center}
\noindent This paper extends the calculation of quantum
corrections to the cosmological correlation
$\langle\zeta\zeta\rangle$, which has been done by Weinberg for a
loop of minimally coupled scalars, to other types of matter loops
and a general and realistic potential.  It is shown here that
departures from scale invariance are never large even when Dirac,
vector, and conformal scalar fields are present \emph{during}
inflation.  No fine tuning is needed, in the sense that effective
masses or coupling constants can have arbitrary values.  Even when
the mass is as large as $M_{Pl}$, the one-loop result is still
naturally smaller than the classical one.  Thus, the results show
that the existence of these fields \emph{during} inflation may not
be ruled out and is consistent with natural reheating.

\vfill

\pagebreak
\setcounter{page}{1}

\begin{center}
{\bf I. INTRODUCTION}
\end{center}
\nopagebreak If we would like to understand how electrons and
photons arise during reheating, it is natural to look back and ask
what happened at the time during inflation.  The gaussian and
nearly scale invariant spectrum predicted by the scalar field
dominated universe theory agrees very well with current
observations.  It is therefore widely believed that the quantum
fluctuations of a scalar field during inflation seeded the large
scale structure of the universe we observe today.  However, in
order to understand where other matter such as fermions and
photons observed today come from, the inflaton\footnote{Although
we do not actually know what the inflaton is and is not, here we
treat any field that has unbroken symmetries, i.e
$\langle\chi\rangle = \langle\psi\rangle = \langle A_{\mu}\rangle
= 0$ as other matter. A scalar field $\varphi$ that has a non-zero
expectation value is considered as an inflaton, as in conventional
belief.  Therefore in the quantum theory of cosmological
fluctuation during inflation considered here, the inflaton
$\varphi$, gravity $g_{\mu\nu}$, and other matter fields $\chi,
\psi, A_{\mu}$ are expanded as $\varphi =
\bar{\varphi}+\delta\varphi, g_{\mu\nu} = \bar{g}_{\mu\nu} +
\delta g_{\mu\nu}, \chi = 0+ \delta\chi,  \psi = 0+ \delta\psi,
A_{\mu}= 0+ \delta A_{\mu}$, respectively.} would have to couple
with these fields during inflation.  There is no reason why there
must be only scalar fields and gravity but nothing else during
inflation.  Other matter would not have arisen during reheating if
there were only a scalar field that coupled exclusively to itself
and gravity.  If other matter such as Dirac, vector, and conformal
scalar fields during inflation do not give anything far larger
than the observed values\footnote{The classical correlation
function is $\langle\zeta\zeta\rangle = \frac{8\pi G H^{2}}{2
(2\pi)^{3}q^{3}|\epsilon|}$} and do not break the scale invariant
of the spectrum in the correlation functions, the existence
of these fields during inflation may not be ruled out. \\

So far, the quantum theory of cosmological fluctuations is
considered up to the quadratic term in the action [1].  Recently,
non-gaussian terms in the scalar field(s) have been calculated
classically [7,8].  The quantum effect to arbitrary order in
cosmological fluctuations has been more recently formulated by
Weinberg[2]. With a sample massless minimally coupled scalar loop
calculation, Weinberg's result shows that the momentum dependence
goes as $q^{-3}\ln q$, with an additional suppression of $GH^{2}$
when compared to the classical result.  Is this true for other
kinds of matter such as Dirac, vector, and conformal scalar fields
as well?  In fact, the unbroken symmetry matter becomes
non-negligible when we go beyond the quadratic term in the action
in cosmological perturbation theories.  It is therefore of great
interest to investigate how the higher-order corrections to the
bilinear correlation function $\langle\zeta\zeta\rangle_{loops}$
depend on momentum $q$ when gravitational fluctuations interact
with general matter other than scalar fields.  Will the result go
approximately as $q^{-3}$ as in the scalar case?\\

If other fields, other than inflaton, acquire a VEV, it is not
necessary that the result will not affect the scale invariance of
the curvature perturbation.  For example, it is considered in the
literature that Dirac fields cannot give a scale invariant
primordial spectrum of density perturbations because their
momentum dependence $\langle\zeta\zeta\rangle$ is far from
$q^{-3}$[6] and vector fields can generate a scale invariant
spectrum only in a special kind of mass[13].  However, the Dirac
and vector fields in [6] and [13] respectively were considered
only \textit{classically}.  In general, Dirac and vector fields in
an expanding universe only exist as quantum fields with zero
expectation value $\langle\psi\rangle = \langle A_{\mu}\rangle =
0$ and what we observe is the density correlation function related
to $\langle\zeta\zeta\rangle$, not the product of the fields
$\langle\delta\bar{\psi}\delta\psi\rangle$, $\langle\delta
A_{i}\delta A_{j}\rangle$.  Therefore, we have to learn how to
quantize such fields at higher order in cosmological fluctuations.
In this paper I calculate the \textit{quantum} effect to the
observable $\langle\zeta\zeta\rangle$ due to other matter loops.
We use the in-in formalism[2,4], appropriate for calculating
expectation values, rather than the S-Matrix in time dependent
backgrounds.  I mainly follow the calculation of Weinberg for a
loop of minimally coupled scalars in [2] and extend it to the
loops of (massive) Dirac, vector, and conformal scalar fields. We
investigate how the $\zeta$ correlation function depends on its
momentum and whether it can be large.
\\

It is also important to investigate the momentum dependence when
quantum corrections are applied to two-point correlation
functions.  If the momentum dependence of the loop spectrum goes
as $q^{-n}$ such that $n$ is far greater than $3$, this will
produce a larger spectrum than the classical value outside the
horizon when $q\rightarrow 0$.  The existence of Dirac and vector
fields during inflation can be easily ruled out if the spectrum is
far from the scale invariant; therefore, those fields cannot be
the candidate for the origin of structure.  We have shown that
this is not the case.  We \textit{always} obtain  nearly scale
invariant spectrums even when Dirac, vector, and conformal scalar
fields are
included.  \\

In section II we explain why quantum effects could be large. In
section III we present the aspects of the in-in formalism and the
un-equal time (anti)commutators of all fluctuations that are
needed for our present purposes.  Section IV and V introduce a
class of theories with a single inflaton field, plus an additional
Dirac field with gravitational interactions.  A (massive) fermion
loop correlation function of $\langle\zeta\zeta\rangle$ is also
calculated in these sections. In sections VI, V II, and VIII we
follow the same approach as in section IV and V except we replace
the Dirac field with a vector field or conformal scalar field
respectively.  In section IX we summarize all the results of this
paper and explain why the departure from scale invariance is still
small even in a more general potential $V(\varphi)\rightarrow
V(\varphi, \bar{\psi}\psi, A_{\mu}A^{\mu})$.  All results show
that for all theories and matter with a general potential
$V(\varphi, \bar{\psi}\psi, A_{\mu}A^{\mu})$, the quantum
correlation functions are never much larger than the classical
(observed) value and are nearly scale invariant.  In appendix we
derive gravitational and general matter fluctuations used in the
loop calculations of section IV-VIII to cubic order for the
general reader.
 \vspace{12pt}
\begin{center}
{\bf II. PROBLEMS }
\end{center}
\nopagebreak
There are some simple arguments that lead us to
believe that quantum effects might contribute to the spectrum in
the order of the observed values without getting suppressed by an
additional factor of $G$.  This could happen when matter couples
with the inflaton and gives a vertex in the order of
$\bar{\varphi}$.  The fact that $\frac{\bar{\varphi}}{M_{Pl}}$ is
\textit{not small }
raises the question whether loop effect could be large.\\

To clarify the problem, we take an example of the interaction of a
fermion $\psi$, inflaton $\varphi$, and graviton $g_{\mu\nu}$ via
\begin{equation}\label{simplearg}
\mathcal{L}_{int} = \sqrt{-g}\varphi\bar{\psi}\psi
\end{equation}
In cosmological fluctuation, we normally expand the fields as
\begin{equation}
g_{\mu\nu} = \bar{g}_{\mu\nu}+\delta g_{\mu\nu},  \varphi =
\bar{\varphi}+ \delta\varphi, \psi = 0+\delta\psi
\end{equation}
In general, fermions interact with fluctuations of both gravity
$\delta g_{\mu\nu}$ and scalar field $\delta \varphi$ and thus
affect the conserved quantity $\zeta$.  However, we can choose a
gauge in which the inflaton does not fluctuate ($\delta\varphi =
0$ gauge [7]) so that $\zeta$ is purely gravity in this gauge.
Therefore, one of the interactions in eq.
 \eqref{simplearg}, for instance the trilinear interaction, has the form
\begin{equation}\label{88}
H_{\zeta\bar{\psi}\psi}(t) = \int d^{3}x
a^{3}(t)\bar{\varphi}(t)\bar{\psi}(\textbf{x},t)\psi(\textbf{x},t)
F[\zeta(\textbf{x},t)]
\end{equation}
where $F[\zeta]$ is some function of $\zeta$, depending on the
details after the expansion of the metric.  Let us choose
$F[\zeta] = \zeta$ for simplicity.  Now we calculate the quantum
contribution of $\langle\zeta\zeta\rangle$\footnote{See the full
formula in the next section or in ref[2]}
\begin{eqnarray}\label{99}
\Big\langle\zeta(\textbf{x},t)\zeta(\textbf{x}',t)\Big\rangle =-
\int_{-\infty}^{t}dt_{2}\int_{-\infty}^{t_{2}}dt_{1}\Big\langle\Big[H_{1},\Big[H_{2},
\zeta(\textbf{x},t)\zeta(\textbf{x}',t)\Big]\Big]\Big\rangle_{0}
\end{eqnarray}
where $H_{1} = H_{I}(t_{1})$, $H_{2} = H_{I}(t_{2})$, and $H_{I}$
is the interaction part of the Hamiltonian (the part that is of
third or higher order in fluctuations) in the interaction picture.
By solving Dirac field equation in an inflating
universe\footnote{This is shown explicitly in fermion sections},
the fermion pair $\bar{\psi}\psi$ goes as $a^{-3}$ at late time.
Therefore, the factor $a^{-3}$ cancels with $\sqrt{-\bar{g}}$ for
each Hamiltonian.  With zero factors of $a(t)$, the result of two
time integrals grows as $(\ln a)^{2}$. But $(\ln a)^{2}$ is not as
significant as $\bar{\varphi}$ in producing an appreciable effect
in the interaction eq. \eqref{88}. Since there are a total of four
$\zeta$s on the RHS of eq. \eqref{99}, we get the factor
$|\zeta^{o}_{q}|^{4}\simeq \frac{(8\pi G H^{2})^{2}}{\epsilon^{2}
q^{6}}$.  The two time integrals become
\begin{equation}
\int dt_{1}\int dt_{2} = \frac{1}{H^{2}}\int
\frac{d\tau_{2}}{\tau_{2}}\int \frac{d\tau_{1}}{\tau_{1}} \simeq
\frac{C}{H^{2}}
\end{equation}
where $\tau$ is the conformal time $\tau \equiv
\int_{t}^{\infty}\frac{dt'}{a(t')}$.  Since $\bar{\varphi}$ does
not change very much during inflation, it can be approximated as
$\bar{\varphi}(t_{1}) \simeq \bar{\varphi}(t_{2})\simeq
\bar{\varphi}(t_{q})$ at the time of horizon exit.  Thus,
$\bar{\varphi}$ does not enter into the time integral. By
collecting the factors of $H,\bar{\varphi}$ and $8\pi G$, we can
then approximate the correlation function as
\begin{equation}
\langle\zeta\zeta\rangle_{loop} \rightarrow \frac{(8\pi G
H^{2}(t_{q}))^{2}\bar{\varphi}^{2}(t_{q})\mathcal{C}_{q}}{H^{2}\epsilon^{2}(t_{q})}
\end{equation}
where $\epsilon \equiv - \frac{\dot{H}}{H^{2}}$ is a slow roll
parameter and $\mathcal{C}_{q}$ is the momentum dependence that
depends on the results of time and momentum integrals and the
details of the propagator to the momentum $p$ and $p'$.  It is
important to note that unlike the term $H$, the term
$\bar{\varphi}$, which could arise via Yukawa coupling,
\textit{does not} give a spectrum that is suppressed by a factor
of $G H^{2}$.  Therefore, if an unperturbed inflaton amplitude is
as large as $M_{Pl}$, we have the correlation function
\begin{equation}
\langle\zeta\zeta\rangle_{loop} \rightarrow \frac{(8\pi
G)H^{2}\mathcal{C}_{q}}{\epsilon^{2}(t_{q})}
\end{equation}
which is in the order of the classical result. \\

The large vertices in the order of the $M_{Pl}$ raise the
possibility that quantum effects are \textit{not} suppressed by a
factor of $G$ as was previously believed.  Therefore the large
vertices could contribute to the loop spectrum in the order of the
classical value. However, such realistic $\bar{\varphi}(t_{q})\sim
M_{Pl}$ coupling can only happen in massive, but not massless,
matter fields at one-loop level.  The reason is that the inflaton
fluctuates around a non-zero background that always contributes to
the non-derivative matter terms in the second order after field
expansions, i.e., $m\bar{\psi}\psi = \bar{\varphi}\bar{\psi}\psi$
or $|\bar{\varphi}|^{2}A_{i}^{2} = m^{2}A_{i}^{2}$.  The massive
case is more general because it allows the possibilities of
interactions with other broken symmetry fields such as the
inflaton and hence gives large vertices in the order of $M_{Pl}$.
Although the argument above is valid, we still need to find out
what $\mathcal{C}_{q}$ is through actual calculations because
$\mathcal{C}_{q}$ is also a function of mass that arises through
the massive propagators.  We have to investigate whether this
$\mathcal{C}_{q}(m)$ gives other kind of suppression.
\vspace{12pt}
\begin{center}
{\bf III. CORRELATION FUNCTION FORMULA}
\end{center}
\nopagebreak
Since the Hamiltonian that governs fluctuations of
fields is explicitly time dependent, we need an in-in
formalism[4].  For the purpose of calculating the in-in
correlation function to arbitrary order, it is convenient to use
Weinberg's formula[2]
\begin{eqnarray}\label{SWInInfor}
\Big\langle Q(t) \Big\rangle &=& \sum_{N =
0}^{\infty}i^{N}\int_{-\infty}^{t}dt_{N}\int_{-\infty}^{t_{N}}dt_{N-1}...
\int_{-\infty}^{t_{2}}dt_{1} \nonumber \\&& \times \Big\langle
\Big[H_{I}(t_{1}),\Big[H_{I}(t_{2}),...\Big[H_{I}(t_{N}),
Q^{I}(t)]...\Big]\Big] \Big\rangle_{0}
\end{eqnarray}
The expectation value on the RHS of the equation above is
annihilated in the free field vacuum by annihilation operators and
on the LHS it is in interacting vacuum.  We will use the equation
above to calculate loop correlation functions. \\

We see from Weinberg's formula above that quantization requires
unequal time (anti) commutators.  The quantity $\zeta(\textbf{x},
t)$, the scalar field $\chi(\textbf{x},t)$, the Dirac field
$\psi(\textbf{x},t)$, and the vector field $A_{i}(\textbf{x},t)$
can be Fourier decomposed as
\begin{eqnarray}
\zeta(\textbf{x}, t) &=& \int d^{3}q
\Big[e^{i\textbf{q}\cdot\textbf{x}}\alpha(\textbf{q})\zeta_{q}(t)+
e^{-i\textbf{q}\cdot\textbf{x}}\alpha^{*}(\textbf{q})\zeta^{*}_{q}(t)\Big]\\
\chi(\textbf{x}, t) &=& \int d^{3}q
\Big[e^{i\textbf{q}\cdot\textbf{x}}a(\textbf{q})\chi_{q}(t)+
e^{-i\textbf{q}\cdot\textbf{x}}a^{*}(\textbf{q})\chi^{*}_{q}(t)\Big]\\
\psi(\textbf{x}, t)&=& \int d^{3}q \sum_{s}
\Big[e^{i\textbf{q}\cdot\textbf{x}}\alpha(\textbf{q},s)U_{\textbf{q},s}(t)+
e^{-i\textbf{q}\cdot\textbf{x}}\beta^{\dag}(\textbf{q},s)V_{\textbf{q},s}(t)\Big]
\\
A_{i}(\textbf{x}, t) &=& \int d^{3}q \sum_{\lambda }
\Big[e^{i\textbf{q}\cdot\textbf{x}}e_{i}(\hat{q},\lambda)
\alpha(\textbf{q},\lambda)\mathcal{A}_{q}(t)+
e^{-i\textbf{q}\cdot\textbf{x}}e^{*}_{i}
(\hat{q},\lambda)\alpha^{*}(\textbf{q},\lambda)\mathcal{A}^{*}_{q}(t)\Big]
\end{eqnarray}
where $s= \pm \frac{1}{2}$ stands for the spin of fermion,
$\lambda = 1, 2$ is a helicity index of a photon, $\lambda = 1, 2,
3$ is a helicity index of a massive vector field.
$e_{i}(\hat{q},\lambda)$ is a polarization vector.  The creation
and annihilation operators satisfy usual (anti) commutation
relations as
\begin{equation}
\Big[\alpha(\textbf{q}),\alpha^{*}(\textbf{q}')\Big] =
\Big[a(\textbf{q}), a^{*}(\textbf{q}')\Big]= \delta^{3}(\textbf{q}
- \textbf{q}')
\end{equation}
\begin{equation}
\Big[\alpha(\textbf{q},\lambda),\alpha^{*}(\textbf{q}',\lambda')\Big]=
\delta_{\lambda\lambda'}\delta^{3}(\textbf{q} - \textbf{q}')
\end{equation}
\begin{equation}
\Big\{\alpha(\textbf{q},s),\alpha^{\dag}(\textbf{q}',s')\Big\} =
\Big\{\beta(\textbf{q},s),\beta^{\dag}(\textbf{q}',s')\Big\}=
\delta_{ss'}\delta^{3}(\textbf{q} - \textbf{q}')
\end{equation}
\begin{equation}
\Big\{\alpha(\textbf{q},s),\alpha(\textbf{q}',s')\Big\} =
\Big\{\beta(\textbf{q},s),\beta(\textbf{q}',s')\Big\}= 0
\end{equation}
\begin{equation}
\Big[\alpha(\textbf{q}),\alpha(\textbf{q}')\Big] =
\Big[a(\textbf{q}), a(\textbf{q}')\Big] =
\Big[\alpha(\textbf{q},\lambda),\alpha(\textbf{q}',\lambda)\Big]=
0
\end{equation}
\\
and $\zeta_{q}(t), \chi_{q}(t), U_{q,s}(t), V_{q,s}(t) $ and
$\mathcal{A}_{q}(t)$ satisfy interaction picture free field
equations in an inflating universe.  With the (anti) commutation
relations above,
the unequal time (anti) commutators of all matter are \\
\begin{equation}
\Big[\zeta(\textbf{x}_{1},t_{1}), \zeta(\textbf{x}_{2},t_{2})\Big]
= 2 i \int d^{3}p e^{i
\textbf{p}\cdot(\textbf{x}_{1}-\textbf{x}_{2})}I m \Big(
\zeta_{\textbf{p}}(t_{1})\zeta^{*}_{\textbf{p}}(t_{2})\Big)
\end{equation}
\begin{equation}
\Big[\chi(\textbf{x}_{1},t_{1}), \chi(\textbf{x}_{2},t_{2})\Big] =
2 i \int d^{3}p e^{i
\textbf{p}\cdot(\textbf{x}_{1}-\textbf{x}_{2})}I m \Big(
\chi_{\textbf{p}}(t_{1})\chi^{*}_{\textbf{p}}(t_{2})\Big)
\end{equation}
\begin{equation}
\Big\{\psi(\textbf{x}_{1},t_{1}),
\bar{\psi}(\textbf{x}_{2},t_{2})\Big\}= \int d^{3}p e^{i
\textbf{p}\cdot(\textbf{x}_{1}-\textbf{x}_{2})}\sum_{s}\Big(
U_{\textbf{p},s}(t_{1})\bar{U}_{\textbf{p},s}(t_{2})+
V_{-\textbf{p},s}(t_{1})\bar{V}_{-\textbf{p},s}(t_{2})\Big)
\end{equation}
and
\begin{equation}
\Big[A_{i}(\textbf{x}_{1},t_{1}),
A_{j}(\textbf{x}_{2},t_{2})\Big]= 2 i\int d^{3}p e^{i
\textbf{p}\cdot(\textbf{x}_{1}-\textbf{x}_{2})}\Pi_{ij}I m\Big(
\mathcal{A}_{\textbf{p}}(t_{1})\mathcal{A}^{*}_{\textbf{p}}(t_{2})\Big)
\end{equation}
where $\Pi_{ij}$ is the polarization factor in which
\begin{equation}\label{lambdamassless}
\Pi_{ij}(\hat{p}) = \sum_{\lambda= 1}^{2}
\hat{e}^{*}_{i}(\hat{p},\lambda)\hat{e}_{j}(\hat{p},\lambda)=
\delta_{ij} - \frac{p_{i}p_{j}}{|\textbf{p}|^{2}}
\end{equation}
is time independent for $m=0$.\\

In this section, we calculate a loop power spectrum of general
matter valid for scalar, fermion and gauge fields.
 To see this in more detail,
 let us consider a general complex bosonic or
 fermionic field $\Psi$ with the interaction\footnote{We can easily generalize to
 realistic interactions (such as the terms containing field derivative) later in
  the next section, once we obtain a general one-loop formula valid for
  any matter in this section.}
\begin{equation}\label{HintGen}
 H_{I}(t) =
\int d^{3}x V(t) \Psi^{*}(\textbf{x},t)\Psi(\textbf{x},t)
\zeta(\textbf{x},t)
\end{equation}
where we omit the spinor and vector indices in this section, and
$V(t)$ is any time dependent vertex function as a consequence of
an expanding universe.  The loop correction to the correlation
function when $N=2$ is
\begin{eqnarray}\label{Q}
\Big\langle Q(t) \Big\rangle =
-\int_{-\infty}^{t}dt_{2}\int_{-\infty}^{t_{2}}dt_{1}\Big\langle\Big[H_{1},\Big[H_{2},
Q \Big]\Big]\Big\rangle_{0}
\end{eqnarray}
If $Q$ is the product of the gravitational field $\zeta$ treated
as the external legs, the vacuum expectation values of matter,
either bosons or fermions, that circulate inside the loop on the
RHS of eq. \eqref{Q} can be evaluated independently from $Q$
because they have different types of creation and annihilation
operators.  We can write a real $\zeta$ field and a complex $\Psi$
field in the interaction picture in terms of creation and
annihilation operators as
\begin{equation}
\zeta(\textbf{x}, t) = \int d^{3}p \Big(e^{i
\textbf{p}\cdot\textbf{x}}\alpha_{\textbf{p}}\zeta_{p}(t)+ e^{-i
\textbf{p}\cdot\textbf{x}}\alpha^{*}_{\textbf{p}}\zeta^{*}_{p}(t)\Big)
\end{equation}
\begin{equation}\label{oppsi}
\Psi(\textbf{x}, t) = \int d^{3}p\sum_{\lambda}\Big( e^{i
\textbf{p}\cdot\textbf{x}}\alpha_{\textbf{p},\lambda}X_{\textbf{p},\lambda}(t)+
 e^{-i
\textbf{p}\cdot\textbf{x}}\beta^{\dag}_{\textbf{p},
\lambda}W_{\textbf{p},\lambda}(t)\Big)
\end{equation}
and
\begin{equation}\label{oppsibar}
\Psi^{*}(\textbf{x}, t) = \int d^{3}p \sum_{\lambda}\Big(e^{-i
\textbf{p}\cdot\textbf{x}}\alpha^{\dag}_
{\textbf{p},\lambda}X^{*}_{\textbf{p},\lambda}(t)+ e^{i
\textbf{p}\cdot\textbf{x}}\beta_{\textbf{p},\lambda}W^{*}_{\textbf{p},\lambda}(t)\Big)
\end{equation}
where $\alpha_{\textbf{p}}$ satisfies the commutation relations
since $\zeta$ is a boson but $\alpha_{\textbf{p},\lambda}$ and
$\beta_{\textbf{p},\lambda}$ satisfy the (anti) commutation
relations for (fermionic) bosonic matter loops.  $\lambda$ is
either the spin index for the fermion or the helicity for the
gauge field.  $X_{\textbf{p},\lambda}(t)$ and
$W_{\textbf{p},\lambda}(t)$ satisfy the matter free field equation
in an inflating universe.
 \\

Let us now evaluate the RHS of eq. \eqref{Q}
\begin{equation}
\Big[H_{2},Q\Big] = \int d^{3}x_{2} V_{2}
\Big[\Psi^{*}_{2}\Psi_{2} \zeta_{2}, Q\Big] = \int d^{3}x_{2}
V_{2} \Psi^{*}_{2}\Psi_{2}(\zeta_{2}Q -Q\zeta_{2})
\end{equation}
\begin{eqnarray}
\Big\langle\Big[H_{1},\Big[H_{2},Q\Big]\Big]\Big\rangle_{0} &=&
\int\int d^{3}x_{1}d^{3}x_{2}V_{1} V_{2}
\Big\langle\Big[\Psi^{*}_{1}\Psi_{1}\zeta_{1},
\Psi^{*}_{2}\Psi_{2}\zeta_{2}Q\Big]\nonumber \\&&-
\Big[\Psi^{*}_{1}\Psi_{1}\zeta_{1},Q
\Psi^{*}_{2}\Psi_{2}\zeta_{2}\Big]\Big\rangle_{0}
\nonumber \\
&=& \int\int d^{3}x_{1}d^{3}x_{2}V_{1} V_{2}
\Big\langle\Big(\Psi^{*}_{1}\Psi_{1}\Psi^{*}_{2}\Psi_{2}(\zeta_{1}
\zeta_{2}Q-\zeta_{1}Q\zeta_{2}) \nonumber
\\&&+\Psi^{*}_{2}\Psi_{2}\Psi^{*}_{1}\Psi_{1} (Q\zeta_{2}
\zeta_{1}- \zeta_{2}Q\zeta_{1}) \Big)\Big\rangle_{0}
\end{eqnarray}
Since $\Psi$ is independent of $\zeta$ and $Q$, the vacuum
expectation value can be evaluated independently.  \\

For the $\zeta$ part,
\begin{equation}
\langle \zeta_{1}\zeta_{2}Q\rangle_{0} = 2\int d^{3}k d^{3}k'
e^{i\textbf{k}\cdot (\textbf{x}_{1}-\textbf{x})+ i
 \textbf{k}'\cdot (\textbf{x}_{2}-\textbf{x}') }
 \zeta_{k}(t_{1})\zeta_{k'}(t_{2})\zeta^{*}_{k}(t)\zeta^{*}_{k'}(t)
\end{equation}
Hence,
\begin{equation}
\langle \zeta_{1}\zeta_{2}Q\rangle _{0}= \langle
Q\zeta_{2}\zeta_{1}\rangle^{*}_{0}
\end{equation}
Similarly,
\begin{equation}
\langle \zeta_{1}Q\zeta_{2}\rangle_{0} = 2\int d^{3}k d^{3}k'
e^{i\textbf{k}\cdot (\textbf{x}_{1}-\textbf{x})+ i
 \textbf{k}'\cdot (\textbf{x}_{2}-\textbf{x}') }
 \zeta_{k}(t_{1})\zeta_{-k'}(t)\zeta^{*}_{k}(t)\zeta^{*}_{-k'}(t_{2})
\end{equation}
Hence,
\begin{equation}
\langle \zeta_{1}Q\zeta_{2}\rangle_{0} = \langle
\zeta_{2}Q\zeta_{1}\rangle^{*}_{0}
\end{equation}
For the matter field, only $\Psi_{1}$ could pair with $\Psi_{2}$
at different times.  To see this in more detail, we write the
field operator $\Psi$ in terms of creation and annihilation
operators similar to that in eq. \eqref{oppsi}. Hence, these
products of the fields can be written in momentum space as
\begin{eqnarray}
\langle \Psi^{*}_{1}\Psi_{1}\Psi^{*}_{2}\Psi_{2} \rangle_{0} =
\int\int d^{3}pd^{3}p' e^{i
(\textbf{p}+\textbf{p}')\cdot(\textbf{x}_{1}-\textbf{x}_{2})}\nonumber
\\\times \sum_{\lambda,\lambda'}
X_{\textbf{p},\lambda}(t_{1})X^{*}_{\textbf{p},\lambda}(t_{2})
W_{\textbf{p}',\lambda'}(t_{2})W^{*}_{\textbf{p}',\lambda'}(t_{1})
\end{eqnarray}
Since
\begin{equation}
\langle \Psi^{*}_{1}\Psi_{1}\Psi^{*}_{2}\Psi_{2}\rangle_{0} =
\langle \Psi^{*}_{2}\Psi_{2}\Psi^{*}_{1}\Psi_{1}\rangle^{*}_{0}
\end{equation}
Hence, the correlation function is
\begin{eqnarray}
\int d^{3}x
e^{i\mathbf{q}\cdot(\mathbf{x}-\mathbf{x}')}\Big\langle
Q(t)\Big\rangle =
-\int_{-\infty}^{t}dt_{2}\int_{-\infty}^{t_{2}}dt_{1}\int
d^{3}x\int d^{3}x_{2}\int d^{3}x_{1} \nonumber \\ \times 2Re \Big(
\langle\Psi^{*}_{1}\Psi_{1}\Psi^{*}_{2}\Psi_{2}\rangle_{0}
(\langle
\zeta_{1}\zeta_{2}Q\rangle_{0}-\langle\zeta_{1}Q\zeta_{2}\rangle_{0})\Big)
\end{eqnarray}
Note that many integrals over the momenta $\textbf{k}$ and
$\textbf{k}'$ can be eliminated via the space integrations that
produce delta functions, i.e.,
\begin{eqnarray}
\int d^{3} x &\rightarrow&
(2\pi)^{3}\delta^{3}(\textbf{q}-\textbf{k})\\
\int d^{3} x_{1} &\rightarrow&
(2\pi)^{3}\delta^{3}(\textbf{q}+\textbf{p}+\textbf{p}')\\
\int d^{3} x_{2} &\rightarrow&
(2\pi)^{3}\delta^{3}(\textbf{k}'-\textbf{p}-\textbf{p}')
\end{eqnarray}
We therefore have the formula
\begin{eqnarray}\label{genformula}
\int d^{3}x e^{i \textbf{q}\cdot
(\textbf{x}-\textbf{x'})}\Big\langle
\zeta(\textbf{x},t)\zeta(\textbf{x}',t)\Big\rangle_{loop} =
-4(2\pi)^{9}\int d^{3}pd^{3}p'
\delta^{3}(\textbf{q}+\textbf{p}+\textbf{p}') \nonumber \\ \times
\int_{-\infty}^{t}dt_{2}V_{2} \int_{-\infty}^{t_{2}}dt_{1}V_{1}
Re\Big(\mathcal{Z}\mathcal{M} \Big)
\end{eqnarray}
where
\begin{eqnarray}\label{ZetaPro}
\mathcal{Z} = \zeta_{q}(t_{1})\zeta^{*}_{q}(t)
\Big(\zeta_{q}(t_{2})\zeta^{*}_{q}(t)-\zeta_{q}(t)\zeta^{*}_{q}(t_{2})\Big)
\end{eqnarray}
and
\begin{eqnarray}\label{matter}
\mathcal{M} = \sum_{\lambda,\lambda'}X_{p,\lambda}(t_{1})
X^{*}_{p,\lambda}(t_{2})W_{p',\lambda'}(t_{2})W^{*}_{p',\lambda'}(t_{1})
\end{eqnarray}
Eq. \eqref{matter} is a formula for a general matter field loop.
\\

For a real scalar field $\chi(\textbf{x}, t) =
\chi^{*}(\textbf{x}, t) $, we have $X_{p}=\chi_{p}$ and $W_{p} =
\chi^{*}_{p}$.  Hence, eq. \eqref{matter} becomes
\begin{eqnarray}\label{mattersigma}
\mathcal{M}_{\chi}= 2\chi_{p}(t_{1})
\chi^{*}_{p}(t_{2})\chi^{*}_{p'}(t_{2})\chi_{p'}(t_{1})
\end{eqnarray}
Note that we have an additional factor of $2$ for any real field.
\\

For the charged scalar field $\chi(\textbf{x},t)\neq
\chi^{*}(\textbf{x},t)$, we still have $X_{p}=\chi_{p}$ and $W_{p}
= \chi^{*}_{p}$.  Hence, eq. \eqref{matter} becomes
\begin{eqnarray}\label{matterchi}
\mathcal{M}_{\chi}= \chi_{p}(t_{1})
\chi^{*}_{p}(t_{2})\chi^{*}_{p'}(t_{2})\chi_{p'}(t_{1})
\end{eqnarray}
which only differs from the real scalar field by a factor of $2$.\\

For a fermionic field which is always complex, we have
$X_{p,\lambda}= U_{p,s}$ and $W_{p,\lambda} = V_{p, s}$. eq.
\eqref{matter} becomes
\begin{eqnarray}\label{matterpsi}
\mathcal{M}_{\psi} = \sum_{r,s= 1,2}U_{p,r}(t_{1})
U^{*}_{p,r}(t_{2})V_{p',s}(t_{2})V^{*}_{p',s}(t_{1})
\end{eqnarray}
For a real vector field $A_{i}(\textbf{x},t) =
A_{i}^{*}(\textbf{x},t)$, we have
$X_{p,\lambda}=\mathcal{A}_{p,\lambda}$ and $W_{p,\lambda} =
\mathcal{A}^{*}_{p,\lambda}$.  Hence, eq.\eqref{matter} becomes
\begin{eqnarray}\label{matterA}
\mathcal{M}_{\mathcal{A}} =
2\sum_{\lambda,\lambda'}\mathcal{A}_{p,\lambda}(t_{1})
\mathcal{A}^{*}_{p,\lambda}(t_{2})\mathcal{A}^{*}_{p',\lambda'}(t_{2})\mathcal{A}_{p',\lambda'}(t_{1})
\end{eqnarray}

Note that the fermion and gauge fields require the summation over
spin and helicity at \textit{different times}, especially in the
massive theories.  In the subsequent sections
 we will use the
formula for the general matter field loop shown above to calculate
the power spectrums for more realistic interactions between
various kinds of matter and gravity.
\vspace{12pt}
\begin{center}
{\bf IV. FERMION LOOP, INFLATON, AND GRAVITY}
\end{center}
\nopagebreak
In the known theory of cosmological fluctuation, only
an inflaton and gravity are considered.  Here we consider
additional fermion as
\begin{eqnarray}\label{totalaction}
\mathcal{L} &=&
\mathcal{L}_{g}+\mathcal{L}_{\varphi}+\mathcal{L}_{f} \nonumber \\
&=& -\frac{1}{2}\sqrt{-g}\Big[\frac{1}{8\pi G}R
+g^{\mu\nu}\partial_{\mu}\varphi\partial_{\nu}\varphi +
2V(\varphi) \nonumber \\&&+
\bar{\psi}\gamma^{\alpha}(\mathcal{D}_{\alpha}\psi) -
(\mathcal{D}_{\alpha}\bar{\psi})\gamma^{\alpha}\psi+ 2 m
\bar{\psi}\psi\Big]
\end{eqnarray}
where $\mu,\nu,...$ are the space time indices, and
$\alpha,\beta,...$ are the Lorentz indices raised and lowered by
the vierbein $V^{\alpha}_{\mu}$.  To deal with fermion, we need
the tetrad formalism[10].  The metric in any general non-inertial
coordinate system is related to the vierbein by
\begin{equation}\label{vv}
g_{\mu\nu}(x) =
V^{\alpha}_{\mu}(x)V^{\beta}_{\nu}(x)\eta_{\alpha\beta}
\end{equation}
The covariant derivative to fermionic field due to the gravity
interaction is
\begin{equation}
\mathcal{D}_{\alpha} \equiv V_{\alpha}^{\mu}\partial_{\mu}+
\frac{1}{2}\sigma^{\beta\gamma}V_{\beta}^{\nu}V_{\alpha}^{\mu}V_{\gamma\nu;\mu}
\end{equation}
where $\sigma^{\beta\gamma}\equiv
\frac{1}{4}[\gamma^{\beta},\gamma^{\gamma}]$.  To find out what
the time dependent propagators are, it is necessary to solve
interaction free field equations in an inflating universe.  For a
fermion, it is
\begin{equation}
a^{-\frac{3}{2}}\frac{d}{dt}\Big(a^{\frac{3}{2}}\gamma^{0}\psi\Big)
+\frac{\gamma^{i}\partial_{i}}{a}\psi + m\psi= 0
\end{equation}
We can re-scale the field $\psi \equiv a^{-\frac{3}{2}}S$ and work
with the conformal time.  We therefore have a simpler Dirac
equation as
\begin{equation}\label{chi}
\gamma^{0}S'+\gamma^{i}\partial_{i}S + m a S = 0
\end{equation}
where $S'$ denotes the conformal time derivative.  Since the
background is spatially translation invariant, the solution can be
written in mode function as
\begin{eqnarray}\label{modepsi}
\psi(\textbf{x}, t)&\equiv& a^{-\frac{3}{2}}(t)S(\textbf{x},
t)\nonumber \\ &=& \frac{1}{a^{\frac{3}{2}}(t)}\int d^{3}q
\sum_{s}
\Big[e^{i\textbf{q}\cdot\textbf{x}}\alpha(\textbf{q},s)u_{\textbf{q},s}(t)+
e^{-i\textbf{q}\cdot\textbf{x}}\beta^{\dag}(\textbf{q},s)v_{\textbf{q},s}(t)\Big]
\end{eqnarray}
where $u_{\textbf{q},s}(t)$ and $v_{\textbf{q},s}(t)$  satisfy
\begin{equation}\label{ufrweq}
\gamma^{0}u'_{\textbf{q},s}+ i\gamma^{i}q_{i}u_{\textbf{q},s} + i
m a u_{\textbf{q},s} = 0
\end{equation}
and
\begin{equation}\label{vfrweq}
\gamma^{0}v'_{\textbf{q},s}- i\gamma^{i}q_{i}v_{\textbf{q},s} + i
m a v_{\textbf{q},s} = 0
\end{equation}
For $\zeta_{q}(t)$, it satisfies Mukhanov's equation [1] as
\footnote{We work with interaction picture free field equations
here to obtain time dependent propagators via their solutions.
When loop effect is included, Mukhanov's equation is modified by
varying the loop quantum effective action with respect to
$\zeta$.}
\begin{equation}\label{zetaq}
\ddot{\zeta}_{q} + \frac{d}{dt}\Big(\ln a^{3}(t)\epsilon\Big)
\dot{\zeta}_{q} +\frac{q^{2}}{a^{2}}\zeta_{q} = 0
\end{equation}
We can expand matter and gravity fluctuations in the actions of
 \eqref{totalaction} to arbitrary order in
cosmological fluctuation.  It is also convenient to write down the
gravitational and all matters actions in ADM form [5] and solve
for $N$ and $N_{i}$ in the constraint equations.  The cubic term
and higher order terms are time dependent vertices that are needed
to calculate loop diagrams.  The direct expansion of matter and
gravitational fluctuations to higher order are complicated.
However, as shown in the appendix, many terms are not necessary
since they are cancelled via field equations and are removed by
the field redefinition of $\zeta$.  Therefore, the important terms
of the trilinear interaction of any general matter are
\begin{equation}\label{1}
H_{\zeta M M}(t) = -\int d^{3}x \epsilon H a^{5}
(T^{00}+a^{2}T^{ii})\nabla^{-2}\dot{\zeta}
\end{equation}
The fermion energy momentum tensor in the presence of gravity is
\begin{equation}\label{3}
T^{\mu\nu}_{f} =
-\frac{1}{2}(\bar{\psi}_{;}^{\mu}\gamma^{\nu}\psi+
\bar{\psi}_{;}^{\nu}\gamma^{\mu}\psi-\bar{\psi}\gamma^{\mu}\psi_{;}^{\nu}-
\bar{\psi}\gamma^{\nu}\psi_{;}^{\mu})
\end{equation}
Therefore, we have the time component of the energy momentum
tensor to quadratic order as
\begin{eqnarray}\label{4}
T^{00}_{f} &=&
\dot{\bar{\psi}}\gamma^{0}\psi-\bar{\psi}\gamma^{0}\dot{\psi} \nonumber \\
&=&(-i)\Big(\dot{\psi}^{\dag}\psi-\psi^{\dag}\dot{\psi} \Big)
\end{eqnarray}
where $\bar{\psi} \equiv \psi^{\dag}\beta$, $\beta \equiv i
\gamma^{0}$, $(\gamma^{0})^{2}= -1$, and $\partial^{0} =
-\partial_{0} = -\frac{\partial}{\partial t}$.  Similarly, the
spatial component of the energy momentum tensor is
\begin{eqnarray}\label{5}
a^{2}T^{ii}_{f} &=&
\frac{1}{a}\Big(\bar{\psi}\gamma^{i}(\partial_{i}\psi)-(\partial_{i}\bar{\psi})\gamma^{i}\psi\Big)
\end{eqnarray}
where $\partial^{i} = \frac{\partial_{i}}{a^{2}}$ and
$\gamma^{\mu} = V^{\mu}_{\alpha}\gamma^{\alpha}$.  Therefore,
$\gamma^{0} = V^{0}_{0}\gamma^{0} = \gamma^{0}$ and $\gamma^{i} =
V^{i}_{m}\gamma^{m} = \frac{\delta^{i}_{m}\gamma^{m}}{a}$.  These
give us the Hamiltonian interaction of fermion and gravity to the
cubic order as
\begin{eqnarray}\label{Hinteraction}
H_{\zeta \bar{\psi}\psi}(t) &=& -\int d^{3}x \epsilon H a^{5}
\Big[\dot{\bar{\psi}}\gamma^{0}\psi-\bar{\psi}\gamma^{0}\dot{\psi}
\nonumber
\\&&+\frac{1}{a}\Big(\bar{\psi}\gamma^{i}(\partial_{i}\psi)
-(\partial_{i}\bar{\psi})\gamma^{i}\psi\Big)
\Big]\nabla^{-2}\dot{\zeta}
\end{eqnarray}
Note that the interaction above is real and also valid for massive
fermions.  The cubic order in the interaction above can be further
simplified with
\begin{eqnarray}
\dot{\bar{\psi}}\gamma^{0}\psi - \bar{\psi}\gamma^{0}\dot{\psi}
+\frac{\partial_{i}\bar{\psi}}{a}\gamma^{i}\psi
-\bar{\psi}\gamma^{i}\frac{\partial_{i}\psi}{a}-2m\bar{\psi}\psi =
0
\end{eqnarray}
Hence, the trilinear interaction Hamiltonian is
\begin{eqnarray}\label{Hint}
H_{\zeta\bar{\psi}\psi}(t) &=& 2\int d^{3}x  \epsilon H
a^{4}\Big(\bar{\psi}\gamma^{i}(\partial_{i}\psi)
-(\partial_{i}\bar{\psi})\gamma^{i}\psi + m
a\bar{\psi}\psi\Big)\nabla^{-2}\dot{\zeta} \nonumber \\
&=& -2\int d^{3}x  \epsilon H a^{5}\Big(
\bar{\psi}\gamma^{0}\dot{\psi}-\dot{\bar{\psi}}\gamma^{0}\psi + m
\bar{\psi}\psi\Big)\nabla^{-2}\dot{\zeta}
\end{eqnarray}
We see that Dirac's equations in an expanding universe
\eqref{ufrweq} and \eqref{vfrweq} at $m=0$ are the same as those
in Minkowski space except that the physical time $t$ is replaced
with the conformal time $\tau$.  So we can expect the plane wave
solutions for $u_{\textbf{q},s}(t) $ and $v_{\textbf{q},s}(t) $ to
be
\begin{equation}
u_{\textbf{q},s}(t) = u_{\textbf{q},s}^{o}e^{-iq\tau},
v_{\textbf{q},s}(t) = v_{\textbf{q},s}^{o}e^{iq\tau}
\end{equation}
where $u_{\textbf{q},s}^{o}$ and $v_{\textbf{q},s}^{o}$ stand for
constant coefficients outside the horizon.  These coefficients can
be determined by matching the solutions at deep inside the horizon
with the flat space solutions.  Deep inside the horizon, the field
does not feel the effect of the expansion of the universe.
Therefore, the normalization factor can be chosen in the same way
as in Minkowski space\footnote{We emphasize that this formula is
\textit{only valid for massless fermion}.  The situation is
entirely different for massive fermion, as will be shown in the
next section.}
\begin{equation}\label{spins}
\sum_{s}u_{q,s}^{o}\bar{u}_{q,s}^{o} =
\sum_{s}v_{q,s}^{o}\bar{v}_{q,s}^{o} =
-\frac{i\gamma^{\mu}q_{\mu}}{2 (2\pi)^{3}q}
\end{equation}
where $q \equiv q^{0} = \sqrt{\textbf{q}^{2}}$.  We see that the
momentum dependence $q$ of the expectation value of the fermion
and anti-fermion pair $\langle\bar{\psi}\psi\rangle$ is far from
the scale invariant spectrum $q^{-3}$.  However, this does not
rule out that fermions could not seed the large scale structure of
the universe observed today.  The reason is that we never observe
the product of either scalar or fermionic fields but rather the
product of temperature $\langle\frac{\delta T}{T}\frac{\delta
T}{T}\rangle$ or density $\langle\frac{\delta
\rho}{\rho}\frac{\delta \rho}{\rho}\rangle$ fluctuations, which
are related to the conserved quantities $\langle
\zeta\zeta\rangle$.  Since fermions interact with the
gravitational fluctuation, we therefore calculate how fermions
affect $\zeta$ at the loop quantum level.  We now calculate the
one-loop graph with two vertices of the two point function.  Owing
to the interaction of gravity and fermionic fields, the quantum
corrections to the $\zeta$ spectrum are
\begin{eqnarray}
\Big\langle\zeta(\textbf{x},t)\zeta(\textbf{x}',t)\Big\rangle_{loop}
&=& -\int_{-\infty}^{t}dt_{2}\int_{-\infty}^{t_{2}}
dt_{1}\Big\langle [H_{1},[H_{2},
\zeta(\textbf{x},t)\zeta(\textbf{x}',t)]\Big\rangle_{0}
\end{eqnarray}
For simplicity, we first start with massless fermions.  We can use
the formula \eqref{genformula} and \eqref{matterpsi} derived in
the previous section.  To match with the interaction Hamiltonian
in eq. \eqref{Hint}, we replace the interaction in eq.
\eqref{HintGen} with
\begin{eqnarray}
\Psi^{*}\Psi &\rightarrow&
\Big(\bar{\psi}\gamma^{i}\frac{\partial_{i}}{a}\psi
-(\frac{\partial_{i}}{a}\bar{\psi})\gamma^{i}\psi\Big)
\\
\zeta_{q}(t_{1,2})&\rightarrow& -\dot{\zeta}_{q}(t_{1,2})/q^{2}\\
V(t) &=& 2\epsilon H a^{5}(t)
\end{eqnarray}
Hence, $\mathcal{Z}$ in eq. \eqref{ZetaPro} changes to
\begin{eqnarray}\label{zeta}
\mathcal{Z} \rightarrow
\frac{1}{q^{4}}\Big(\dot{\zeta}_{q}(t_{1})\zeta^{*}_{q}(t)
(\dot{\zeta}_{q}(t_{2})\zeta^{*}_{q}(t)-\zeta_{q}(t)\dot{\zeta}^{*}_{q}(t_{2}))\Big)
\end{eqnarray}
and $\mathcal{M}_{\psi}$ in eq. \eqref{matterpsi} changes to
\begin{eqnarray}\label{F}
\mathcal{M}_{\psi} \rightarrow
-\frac{1}{a_{1}a_{2}}(p_{i}-p'_{i})(p_{j}-p'_{j})\sum_{r,s}
\gamma^{i}U_{\textbf{p},s}(t_{1})
\bar{U}_{\textbf{p},s}(t_{2})\gamma^{j}V_{\textbf{p}',r}(t_{2})\bar{V}_{\textbf{p}',r}(t_{1})
\end{eqnarray}
The equation above shows the need to sum over spins at different
times.  Fortunately, massless fermions are conformally flat so we
can still use the spin sum formula from flat space.  As seen from
Dirac's equation, the solutions of massless fermions are just
plane waves with conformal time, $u_{\textbf{p},s}(\tau) =
u_{\textbf{p},s}^{o}e^{-i p\tau}$, after re-scaling the field such
that $U_{\textbf{p},s} = a^{-\frac{3}{2}}(t)u_{\textbf{p},s}(t)$.
With the spin sum equation  \eqref{spins}, eq. \eqref{F} becomes
\begin{eqnarray}\label{FF}
\mathcal{M}_{\psi} =
(p_{i}-p'_{i})(p_{j}-p'_{j})\frac{p_{\alpha}p'_{\beta}}{4(2\pi)^{6}pp'a_{1}^{4}a_{2}^{4}}tr(
\gamma^{i}\gamma^{\alpha}\gamma^{j}\gamma^{\beta})e^{-i(p+p')(\tau_{1}-\tau_{2})}
\end{eqnarray}
Since
\begin{equation}
tr( \gamma^{i}\gamma^{\alpha}\gamma^{j}\gamma^{\beta}) =
4(\eta^{i\alpha}\eta^{j\beta}-\eta^{ij}\eta^{\alpha\beta}+\eta^{i\beta}\eta^{\alpha
j})
\end{equation}
we have,
\begin{eqnarray}\label{FFF}
\mathcal{M}_{\psi} =
\frac{1}{(2\pi)^{6}a_{1}^{4}a_{2}^{4}}(p-p')^{2}(1+\hat{p}\cdot
\hat{p}')e^{-i(p+p')(\tau_{1}-\tau_{2})}
\end{eqnarray}
Substituting $\mathcal{M}_{\psi}$ into eq. \eqref{genformula} with
$V(t) = 2\epsilon H a^{5}(t)$, we have
\begin{eqnarray}\label{fermimasslessresult}
\int d^{3}x e^{i \textbf{q}\cdot
(\textbf{x}-\textbf{x'})}\Big\langle
\zeta(\textbf{x},t)\zeta(\textbf{x}',t)\Big\rangle_{loop} =
-16(2\pi)^{3}\int
d^{3}pd^{3}p' \delta^{3}(\textbf{q}+\textbf{p}+\textbf{p}') \nonumber \\
(p-p')^{2}(1+\hat{p}\cdot\hat{p}')
\int_{-\infty}^{t}dt_{2}\epsilon_{2}H_{2}a_{2}
\int_{-\infty}^{t_{2}}dt_{1}\epsilon_{1}H_{1}a_{1}
 Re\Big(\mathcal{Z} e^{-i(p+p')(\tau_{1}-\tau_{2})}\Big)
\end{eqnarray}
where $\mathcal{Z}$ is the contribution of the $\zeta$ part in eq.
\eqref{zeta}.  To calculate $\mathcal{Z}$, we use the solution of
the free field Mukhanov's equation in the interaction picture
\begin{equation}
\zeta_{q}(t) = \zeta^{o}_{q}e^{-i q\tau}(1+i q\tau)
\end{equation}
where
\begin{equation}\label{coeffzeta}
|\zeta^{o}_{q}|^{2} = \frac{8\pi G
H^{2}(t_{q})}{2(2\pi)^{3}\epsilon(t_{q})q^{3}}
\end{equation}
Hence,
\begin{equation}\label{zetapart}
\mathcal{Z} =
\frac{|\zeta^{o}_{q}|^{4}}{H_{1}H_{2}a_{1}^{2}a_{2}^{2}}\Big(
e^{2i q\tau -iq(\tau_{1}+\tau_{2})}-e^{i q
(\tau_{2}-\tau_{1})}\Big)
\end{equation}
During slow roll inflation, we approximate
$\epsilon_{1}\approx\epsilon_{2}\approx\epsilon(t_{q})$.
Integrating over conformal time $\tau_{1}$, we get
\begin{eqnarray}\label{formlt1}
\int d^{3}x e^{i \textbf{q}\cdot
(\textbf{x}-\textbf{x}')}\Big\langle
\zeta(\textbf{x},t)\zeta(\textbf{x}',t)\Big\rangle_{loop} =
-16(2\pi)^{3}|\zeta_{q}^{o}|^{4}\epsilon^{2}(t_{q})\int
d^{3}pd^{3}p' \nonumber
\\\times\delta^{3}(\textbf{q}+\textbf{p}+\textbf{p}')
(p-p')^{2}(1+\hat{p}\cdot\hat{p}')
Re\int_{-\infty}^{0}d\tau_{2}\frac{i}{q+p+p'}(e^{-2iq\tau_{2}}-1)
\end{eqnarray}
where an upper limit $t\rightarrow\infty$ or $\tau\rightarrow 0$
means a time still during inflation but sufficiently late so that
$a(t)$ is many e-foldings larger than its value when $\frac{q}{a}$
falls below $H$.  Integrating over conformal time $\tau_{2}$ gives
\begin{eqnarray}\label{inttau2}
Re\int_{-\infty}^{0}d\tau_{2}\frac{i}{q+p+p'}(e^{-2iq\tau_{2}}-1)
= -\frac{1}{2 q (q+p+ p')}
\end{eqnarray}
Substituting  eqs. \eqref{coeffzeta} and \eqref{inttau2} into eq.
\eqref{formlt1}, we have
\begin{eqnarray}\label{formlp}
\int d^{3}x e^{i \textbf{q}\cdot
(\textbf{x}-\textbf{x}')}\Big\langle
\zeta(\textbf{x},t)\zeta(\textbf{x}',t)\Big\rangle_{loop} = \frac{
2(8\pi G H^{2}(t_{q}))^{2}}{(2\pi)^{3} q^{7}}\nonumber\\\times\int
d^{3}pd^{3}p' \delta^{3}(\textbf{q}+\textbf{p}+\textbf{p}')
\frac{(p-p')^{2}}{q+p+p'}(1+\hat{p}\cdot\hat{p}')
\end{eqnarray}
Power counting shows that the results of the momentum integrals
$p$ ,$p'$ will go as $q^{4}$.  If we use dimensional
regularization to remove UV divergences, the finite part of
$\langle \zeta\zeta\rangle$ for the massless fermion loop will go
as $q^{-3}\ln q$.  To determine the coefficient of the finite part
of the momentum integral above, we will follow the calculation as
done in [2] for the scalar case.  Note that, in general,
\begin{eqnarray}\label{genp}
\int d^{3}p \int d^{3}p'
\delta^{3}(\textbf{p}+\textbf{p}'+\textbf{q})f(p,p',q) = \int
d^{3}p f(p, p'= |\textbf{p}'|= |\textbf{p}+\textbf{q}|, q)
\end{eqnarray}
and
\begin{eqnarray}\label{sph}
\int d^{3}p = \int_{0}^{\infty}p^{2} d p \int_{-1}^{1}d(cos
\theta) \int_{0}^{2\pi}d \varphi
\end{eqnarray}
The conservation of momentum $p' = |\textbf{p}+\textbf{q}|$ gives
\begin{eqnarray}
p'^{2} = p^{2} + q^{2} + 2 p q \cos\theta_{pq}
\end{eqnarray}
where $\theta_{pq}$ is an angle between the vectors $\textbf{p}$
and $\textbf{q}$.  Since $q$ is a fixed external momentum, we can
choose in the $z-$ direction.  Hence, $\theta_{pq} = \theta$ and
\begin{eqnarray}\label{pprime}
p' d p' = p q d (\cos \theta)
\end{eqnarray}
With eqs. \eqref{genp}, \eqref{sph}, and  \eqref{pprime}, we have
\begin{eqnarray}
\int d^{3}p \int d^{3}p'
\delta^{3}(\textbf{p}+\textbf{p}'+\textbf{q})f(p,p',q) =
\frac{2\pi}{q}\int_{0}^{\infty}p d p \int_{|p-q|}^{|p+q|} p' d p'
f(p,p',q)
\end{eqnarray}
Since $2 \textbf{p}\cdot \textbf{p}' = q^{2}-p^{2}-p'^{2}$,  eq.
\eqref{formlp} can be written as
\begin{eqnarray}\label{mldimp}
\int d^{3}x e^{i \textbf{q}\cdot
(\textbf{x}-\textbf{x}')}\Big\langle
\zeta(\textbf{x},t)\zeta(\textbf{x}',t)\Big\rangle_{loop} =
\frac{2 (8\pi G
H^{2}(t_{q}))^{2}}{(2\pi)^{3}q^{7}}\Big[\frac{2\pi}{q}\mathcal{J}(q)\Big]
\end{eqnarray}
where
\begin{eqnarray}\label{Jint}
\mathcal{J}(q) \equiv \int_{0}^{\infty}p d p \int_{|p-q|}^{|p+q|}
p' d p' \frac{(p-p')^{2}}{q+p+p'}\Big(1+
\frac{q^{2}-p^{2}-p'^{2}}{2 p p'}\Big)
\end{eqnarray}
With dimensional regularization, the UV divergence of the integral
above for $\delta = 0$ gives the pole term as
\begin{eqnarray}
\frac{2\pi}{q}\mathcal{J}(q) \Rightarrow q^{4+\delta}F(\delta) =
q^{4}e^{\delta \ln q} F(\delta)
\end{eqnarray}
where
\begin{eqnarray}
F(\delta)\rightarrow\frac{F_{0}}{\delta}+ F_{1}
\end{eqnarray}
Therefore, in the limit $\delta \rightarrow 0$,
\begin{eqnarray}
\frac{2\pi}{q}\mathcal{J}(q) = q^{4}\Big( \frac{F_{0}}{\delta}
+F_{0}\ln q + F_{1} \Big) = q^{4}\Big[F_{0}\ln q + L\Big]
\end{eqnarray}
where $L$ is a divergent constant.  To calculate the coefficient
$F_{0}$, it is necessary to evaluate the integral \eqref{Jint}.
The integral over $p'$ gives\footnote{We use Mathematica for the
integrals and six derivatives.}
\begin{eqnarray}\label{Jprime}
\mathcal{J}(q) = \frac{1}{2}\int_{0}^{\infty} d p \Big(
p(15p^{2}+17 p q +4 q^{2})p'-\frac{pp'^{2}}{2}(11p+6 q)\nonumber
\\+\frac{p'^{3}}{3}(5 p + q) -\frac{p'^{4}}{4}-
4p(p+q)(2p+q)^{2}\log[p+q+p'] \Big)|_{p'=|p-q|}^{p'=p+q}
\end{eqnarray}
To eliminate the divergence in the momentum integral above, we
differentiate $\mathcal{J}(q)$ in eq.  \eqref{Jprime} six times
and then do the integration over $p$.  This gives a finite result
as
\begin{eqnarray}
\frac{d^{(6)}\mathcal{J}(q)}{d q^{6}} &=& - \frac{8}{q}
\end{eqnarray}
where we take the limit $q\rightarrow 0$ \textit{after}
integrating over $p$.  Hence,
\begin{eqnarray}
\mathcal{J}(q) = q^{5}\Big(-\frac{8}{5!}\ln q + L\Big)
\end{eqnarray}
or
\begin{equation}
 F_{0} = -\frac{2\pi}{15}
\end{equation}
Substituting $\mathcal{J}(q)$ back into eq. \eqref{mldimp}, we
have the finite part of correlation function as
\begin{eqnarray}\label{mlcorr}
\int d^{3}x e^{i \textbf{q}\cdot
(\textbf{x}-\textbf{x}')}\Big\langle
\zeta(\textbf{x},t)\zeta(\textbf{x}',t)\Big\rangle_{loop} =-
\frac{ 4\pi (8\pi G H^{2}(t_{q}))^{2}}{ 15(2\pi)^{3}
q^{3}}\Big[\ln q + C\Big]
\end{eqnarray}
with $C$, an unknown constant.  Notice that we have the
\textit{same} sign as that in the massless scalar loop because we
do not have the time ordered product of fermion pairs in eq.
\eqref{SWInInfor}.  The opposite sign of fermion loops only arises
in the in-out theory when we time order the product of fermion
pairs in order to close the loop.  Moreover, the result in eq.
\eqref{mlcorr} is smaller than the classical result by a factor of
$G H^{2}$.
\vspace{12pt}
\begin{center}
{\bf V. MASSIVE FERMION}
\end{center}
\nopagebreak
The calculation is much more difficult for massive
fermions. This is because the mode solution of a massive fermion
at arbitrary wavelength during inflation is not a simple plane
wave like in the massless case or in flat space.  We cannot rely
on the trace technology normally used for spinors in flat space
since the spin sum at different times cannot be written in the
compact form of $\gamma$ matrices.  We therefore need to solve the
massive Dirac equation in an expanding universe during inflation
and perform the spin sum at
different times by multiplying the matrices. \\

We define $u_{s,\textbf{p}}(\tau)\equiv[u_{+,\textbf{p}}(\tau)S,
u_{-,\textbf{p}}(\tau)S]^{T}$ and
$v_{s,\textbf{p}}(\tau)\equiv[v_{+,\textbf{p}}(\tau)S,
v_{-,\textbf{p}}(\tau)S]^{T}$ where $S$ are the two component
eigenvectors of the helicity operators.  We use the Dirac
representation of the gamma matrices.
\begin{eqnarray}\label{gamma1}
    \gamma^{0}&=&(-i)\left(%
\begin{array}{cc}
  1 & 0 \\
  0 & -1 \\
\end{array}%
\right)
\\ \gamma^{i}&=& (-i)\left(%
\begin{array}{cc}
  0 & \sigma^{i}\\
  -\sigma^{i} & 0 \\
\end{array}%
\right)
\end{eqnarray}
Therefore, eq. \eqref{ufrweq} gives
\begin{equation}
u'_{\pm,\textbf{p}} + i (\vec{\sigma}\cdot \vec{p})
u_{\mp,\textbf{p}} \pm im a u_{\pm,\textbf{p}} = 0
\end{equation}
We see that the equations above are first order coupled
differential equations.  To decouple them, we differentiate those
equations one more time.  With some algebra, we get two uncoupled
second-order differential equations,
\begin{equation}
u''_{\pm,\textbf{p}}+ (\vec{p}^{2}+(ma)^{2}\pm
i(am)')u_{\pm,\textbf{p}} = 0
\end{equation}
where $(\vec{\sigma}\cdot \vec{p})^{2} =
(p_{x}^{2}+p_{y}^{2}+p_{z}^{2})\left(%
\begin{array}{cc}
  1 & 0\\
   0 & 1 \\
\end{array}%
\right)=\vec{p}^{2}1_{2\times 2} \equiv p^{2}$.  Therefore, the
equations above are solvable as
\begin{equation}\label{u}
u''_{\pm, \textbf{p}}+ \Big(\vec{p}^{2}+\frac{1}{\tau^{2}}(
r^{2}\pm i r)\Big)u_{\pm,\textbf{p}}=0
\end{equation}
The solutions are
\begin{eqnarray}\label{ufrw}
u_{FRW}(\textbf{p},s,t) &\equiv& \left(%
\begin{array}{cc}
  u_{+,\textbf{p}}\times S \\
  u_{-,\textbf{p}} \times S \\
\end{array}%
\right) \equiv
\left(%
\begin{array}{cc}
  u_{\mu,\textbf{p}}\times S \\
  (\hat{p}\cdot\vec{\sigma})u_{\bar{\mu},\textbf{p}}\times S \\
\end{array}%
\right)\nonumber \\ &=&
\left(%
\begin{array}{cc}
   c_{+,p}\sqrt{-\tau}H_{\mu}^{(1)}(-p\tau)\times S\\
   (\hat{p}\cdot\vec{\sigma})c_{-,p}\sqrt{-\tau}H_{\bar{\mu}}^{(1)}(-p\tau)\times S \\
\end{array}%
\right)
\end{eqnarray}
where $\mu \equiv \frac{1}{2}-i r$, $\bar{\mu} \equiv
\frac{1}{2}+ i r$, and $r \equiv \frac{m}{H}$.\\

We choose the initial conditions so that the positive frequency
mode solutions match with the flat space-time solutions deep
inside the horizon [15]\footnote{Note that the asymptotic behavior
of Hankel's function for large $x \equiv -p\tau$ is
$H_{\nu}^{(1)}(x)\rightarrow \sqrt{\frac{2}{\pi x}}e^{i x -
i\nu\pi/2 -i\pi/4}(1+\frac{i}{2 x}(\nu+\frac{1}{2})(\nu
-\frac{1}{2})+...)$}
\begin{equation}\label{uflat}
u_{flat} = \Big( \frac{E + m}{2 (2\pi)^{3}E}\Big)^{\frac{1}{2}}\left(%
\begin{array}{cc}
  1\times S \\
  \frac{\vec{p}\cdot\vec{\sigma}/a}{E+m} \times S\\
\end{array}%
\right)e^{i\int_{t}^{\infty} E(t') dt'}
\end{equation}
where $E^{2}\equiv m^{2}+ \frac{\vec{p}^{2}}{a^{2}}$, $S =\left(%
\begin{array}{cc}
  1 \\
  0\\
\end{array}%
\right) $ for spin up and
$S =\left(%
\begin{array}{cc}
  0 \\
  1\\
\end{array}%
\right) $ for spin down.  Note that $S^{\dag}S = 1$.  We can check
that, in the massless limit,
$u_{flat} = \frac{1}{\sqrt{2(2\pi)^{3}}}e^{-i p \tau}$, as expected. \\

To find the normalization coefficients $c_{\pm}(p)$, we match the
solutions of eqs. \eqref{ufrw} with  \eqref{uflat} by using the
asymptotic property of Hankel's functions.  Hence,
\begin{equation}
c_{\pm,p} = \frac{i\sqrt{ \pi
p}}{2(2\pi)^{\frac{3}{2}}}e^{\pm\frac{ \pi r }{2 }}
\end{equation}
To calculate $\mathcal{M}_{\psi}$, we use the formula
\eqref{genformula} and \eqref{matterpsi} derived in the previous
section.  To take account of the more realistic interaction
Hamiltonian that arises after the expansion in eq. \eqref{Hint},
we replace the interaction in eq.
  \eqref{HintGen}  with
\begin{eqnarray}
\Psi^{*}\Psi &\rightarrow&
\Big(\bar{\psi}\gamma^{0}\dot{\psi}-\dot{\bar{\psi}}\gamma^{0}\psi
+ m \bar{\psi}\psi\Big)
\\
\zeta_{q}(t_{1,2})&\rightarrow& -\dot{\zeta}_{q}(t_{1,2})/q^{2}\\
V(t) &=& -2\epsilon H a^{5}(t)
\end{eqnarray}
$\mathcal{Z}$ is still the same as that in eq. \eqref{zeta}.
However, $\mathcal{M}_{\psi}$ for massive fermion in eq.
\eqref{matterpsi} becomes
\begin{eqnarray}\label{massfresult}
\mathcal{M}_{\psi}
&=&\frac{2}{a_{1}^{3}a_{2}^{3}}(\hat{p}\sigma^{*}_{2}+
\hat{p}'\bar{\sigma}^{*}_{2})\cdot(\hat{p}\sigma_{1}+
\hat{p}'\bar{\sigma}_{1})\nonumber \\
 &\equiv& \frac{2}{a_{1}^{3}a_{2}^{3}}\vec{\pi}^{*}_{2}\cdot
\vec{\pi}_{1}
\end{eqnarray}
where the $\sigma$s are
\begin{eqnarray}\label{sigmaresult}
\sigma &=&
u_{\bar{\mu},p'}\dot{u}_{\mu,p}-\dot{u}_{\bar{\mu},p'}u_{\mu,p}+ i
m u_{\bar{\mu},p'}u_{\mu,p}
\end{eqnarray}
\begin{eqnarray}\label{sigmabarresult}
\bar{\sigma} &=& \sigma(m \rightarrow -m)
\end{eqnarray}
and $u_{\mu,\bar{\mu}}$ are the massive mode solutions of Dirac
equations
\begin{eqnarray}\label{modeumumubar}
u_{\mu}(x) &=& \frac{i\sqrt{\pi
x}}{2(2\pi)^{\frac{3}{2}}}e^{\frac{\pi
r}{2}}H^{(1)}_{\mu}(x)\\
u_{\bar{\mu}}(x) &=& \frac{i\sqrt{\pi
x}}{2(2\pi)^{\frac{3}{2}}}e^{\frac{-\pi
r}{2}}H^{(1)}_{\bar{\mu}}(x)
\end{eqnarray}
for $x= -p \tau$.  The factor $e^{\pm \frac{\pi r}{2 }}$ arises
due to the fixing of the coefficients $c_{\pm, p}$ of mode
solutions deep inside the horizon with the flat space-time
solutions. We can check that in the massless limit $m = 0$,
$u_{\mu}= u_{\bar{\mu}} = u_{\frac{1}{2}} =
\frac{1}{\sqrt{2}(2\pi)^{\frac{3}{2}}}e^{-ip\tau}$ and $\sigma(t)
= \bar{\sigma}(t) = \frac{i}{2(2\pi)^{3}a}(p'-p)e^{-i(p+p')\tau}$.
Hence,
\begin{eqnarray}\label{fmlresult}
\mathcal{M}|_{m=0} &=&
\frac{4}{a_{1}^{3}a_{2}^{3}}\sigma^{*}_{2}\sigma_{1}(1+
\hat{p}\cdot\hat{p}')\nonumber \\
&=& \frac{1}{(2\pi)^{6}a_{1}^{4}a_{2}^{4}}(p-p')^{2}(1+
\hat{p}\cdot\hat{p}')e^{-i(p+p')(\tau_{1}-\tau_{2})}
\end{eqnarray}
which agrees with the result in massless fermion section in eq. \eqref{FFF}.  \\
Substituting eqs. \eqref{coeffzeta} and \eqref{zetapart} back into
the power spectrum formula eq. \eqref{genformula} with $V(t) = -2
\epsilon H a^{5}(t)$, we have $\langle\zeta\zeta\rangle$ due to
massive fermion loop as
\begin{eqnarray}\label{for10}
\int d^{3}x e^{i \textbf{q}\cdot
(\textbf{x}-\textbf{x}')}\Big\langle
\zeta(\textbf{x},t)\zeta(\textbf{x}',t)\Big\rangle_{loop} =
-\frac{4(2\pi)^{3}(8\pi G H^{2})^{2}}{q^{6}}\nonumber \\\times
\int d^{3}pd^{3}p'
\delta^{3}(\textbf{q}+\textbf{p}+\textbf{p}')\mathcal{I}
\end{eqnarray}
where
\begin{eqnarray}\label{I33}
\mathcal{I} = Re \int_{-\infty}^{\infty}d t_{2}a_{2}^{3} (
e^{-iq\tau_{2}}-e^{i q \tau_{2}})
\int_{-\infty}^{t_{2}}dt_{1}a_{1}^{3}e^{-iq\tau_{1}}\mathcal{M}_{\psi}
\end{eqnarray}
where $\mathcal{M}_{\psi}$ is the result contributed by the
fermionic part in eq. \eqref{massfresult} and the exponential
terms $e^{\pm i q\tau}$ in eq. \eqref{I33} come from the
$\zeta_{q}$ parts.
\\

So far, the result in eq.  \eqref{massfresult} is exact.  There is
no approximation involved in eq.  \eqref{massfresult}.  However,
the exact result involves the integrand of Hankel's functions and
their time derivatives and this makes the integration quite
challenging.  Nevertheless, we get some idea that the time
integrals will converge and go at most as $(\log a)^{2}$. However,
in order to determine the momentum dependence of the power
spectrum, we need to integrate over unequal times $t_{1}, t_{2}$
and momentums $p,p'$ associated with fermion fluctuations. We will
first integrate over times and then over momentums.  The direct
calculation is complicated because it involves integrating over
products of Hankel's functions with complex order.  Since $p$ is
the running momentum from $0$ to $\infty$ whereas $q$ is the fixed
external momentum associated with a conserved quantity
$\zeta_{q}$, it is helpful to divide the integrals over momentum
$p$ in eq. \eqref{for10} as an integral when $\Lambda q \leq p
\leq \infty$ and $0 \leq p \leq \Lambda q$.  The first integral
when $p \rightarrow \infty$ can be approximated as if the fermion
is massless due to its high momentum.  So, the result at high
momentum after dimensional regularization will be close to that in
the massless case in eq. \eqref{mlcorr}.  The second integral,
when $p \leq  \Lambda q$, indicates that the mass effect may
become important in the result.  Therefore, additional
calculations are needed to determine the
momentum dependence.\\

The momentum $p$ corresponds to the fermion fluctuation $\psi_{p}$
and the momentum $q$ corresponds to the conserved quantity
$\zeta_{q}$ fluctuation, this implies that a massive fermion
$\psi_{p}$ exits the horizon \textit{before} $\zeta_{q}$ exits the
horizon if $\Lambda$ is in the order of $1$.  In other words, the
fermion fluctuation $a^{3}\bar{\psi}_{p'}\psi_{p}$ is frozen by
the time the fluctuation $\zeta_{q}$ crosses the horizon.  Outside
the horizon of $\zeta_{q}$, $p,p'$ are sufficiently small and
Hankel's function can be approximated as [11]
\begin{eqnarray}\label{approx}
H^{(1)}_{\beta}(x) \approx
-\frac{i\Gamma(\beta)}{\pi}\Big(\frac{x}{2} \Big)^{-\beta}
\end{eqnarray}
which is valid for $x \ll 1$ and $\beta > 0$.  Hence, the mode
solutions in eq.  \eqref{modeumumubar} become
\begin{eqnarray}
u_{\mu}^{o}(x) &=& \frac{\Gamma(\mu) e^{\frac{\pi r}{2}}}
{(2\pi)^{2}}(\frac{x}{2})^{i r}\\
u_{\bar{\mu}}^{o}(x) &=& \frac{\Gamma(\bar{\mu}) e^{\frac{-\pi
r}{2}}}{(2\pi)^{2}}(\frac{x}{2})^{-i r}
\end{eqnarray}
With this approximation, their (conformal) time derivatives are
\begin{equation}
u_{\mu}^{'o} = \frac{i r}{\tau}u_{\mu}^{o},  u_{\bar{\mu}}^{'o} =
-\frac{i r}{\tau}u^{o}_{\bar{\mu}}
\end{equation}
Hence,
\begin{eqnarray}\label{infra}
\sigma^{o} &=& -i m u_{\bar{\mu},p'}^{o}u_{\mu, p}^{o}\\
u_{\bar{\mu},p'}^{o}u_{\mu,p}^{o}&=& \frac{|\Gamma(\mu)|^{2}}{(2\pi)^{4}}p^{ir}p'^{-ir}\\
\sigma^{*o}_{2}\sigma_{1}^{o} =
\bar{\sigma}^{*o}_{2}\bar{\sigma}_{1}^{o}&=&
\frac{m^{2}|\Gamma(\mu)|^{4}}{(2\pi)^{8}}\\
\sigma^{*o}_{2}\bar{\sigma}_{1}^{o} =
(\bar{\sigma}^{*o}_{2}\sigma_{1}^{o})^{*}&=&
-\frac{m^{2}|\Gamma(\mu)|^{4}}{(2 \pi)^{8}}p'^{2 i r}p^{-2 i r}
\end{eqnarray}
Notice that $\sigma(p,p')$ and $\bar{\sigma}(p,p')$ are both
\textit{time independent}.  The exponential factors $e^{\pm \pi
r}$ arise when we fix the coefficients to match the solutions with
those inside the horizon.  Hence, from eq.  \eqref{massfresult}
\begin{eqnarray}\label{Fpp}
\mathcal{M}^{o}_{\psi} &=& \frac{2
m^{2}|\Gamma(\mu)|^{4}}{(2\pi)^{8}a_{1}^{3}a_{2}^{3}}\Big(
2- (\hat{p}\cdot\hat{p}')(p'^{2 i r}p^{-2ir}+ c.c.) \Big) \nonumber \\
&\equiv& a_{1}^{-3}a_{2}^{-3}\mathcal{F}^{o}_{p,p'}
\end{eqnarray}
where the factors $a_{1,2}^{-3}$ from the fermionic part will
cancel with the factor $\sqrt{-\bar{g}}$ in eq. \eqref{I33}.
Therefore,
\begin{eqnarray}\label{J}
\mathcal{I}^{o} &=& \mathcal{F}^{o}_{p,p'}Re
\int_{-\infty}^{\infty} d t_{2} (e^{-iq\tau_{2}}-e^{i q \tau_{2}})
\int_{-\infty}^{t_{2}}d t_{1} e^{-iq\tau_{1}}\nonumber \\
&=&\frac{\mathcal{F}^{o}_{p,p'}}{H^{2}(t_{q})}Re
\int_{-\infty}^{0} \frac{d\tau_{2}}{\tau_{2}} (e^{
-iq\tau_{2}}-e^{i q \tau_{2}})
\int_{-\infty}^{\tau_{2}}\frac{d\tau_{1}}{\tau_{1}}e^{-iq\tau_{1}}
\end{eqnarray}
Although we see from eq. \eqref{J} that the time integral is of
the order of $(\log a)^{2}$, we need to evaluate this integral if
we want to see the momentum dependence $q$ for the correlation
function of $\zeta$.
\\
From eq.  \eqref{J}, we have
\begin{eqnarray}\label{JJ}
\mathcal{I}^{o} &=&\frac{\mathcal{F}^{o}_{p,p'}}{H^{2}(t_{q})}Re
\int_{-\infty}^{0} \frac{d\tau_{2}}{\tau_{2}} (e^{
-iq\tau_{2}}-e^{i q \tau_{2}})Ei(-i q \tau_{2})
\end{eqnarray}
Using $Ei(-i x) = ci(x)- i si(x)$, we have
\begin{eqnarray}\label{JJJ}
\mathcal{I}^{o}
&=&-\frac{2\mathcal{F}^{o}_{p,p'}}{H^{2}(t_{q})}\int_{-\infty}^{0}
\frac{d\tau_{2}}{\tau_{2}} \sin(q\tau_{2}) si (q\tau_{2})
\end{eqnarray}
With Mathematica,
\begin{eqnarray}
\int_{-\infty}^{0}\frac{d\tau_{2}}{\tau_{2}}\sin(q\tau_{2})si(q\tau_{2})
= -\frac{\pi^{2}}{8}
\end{eqnarray}
Therefore,
\begin{eqnarray}\label{JJJJ}
\mathcal{I}^{o} &=&\frac{\pi^{2}\mathcal{F}^{o}_{p,p'}}{4 H^{2}}
\end{eqnarray}
Notice that the result of the time integral above is
$q$-independent.  Hence,
\begin{eqnarray}\label{deri}
\int_{|p-q|}^{p+q} p' d p' \mathcal{F}^{o}_{p,p'}=
G^{o}(p+q)-G^{o}(|p-q|) = 2q \frac{\partial G^{o}}{\partial p} +
\mathcal{O}(q^{3})
\end{eqnarray}
Note that we only keep the leading order in $q$ in the last
equation.  Hence,
\begin{eqnarray}\label{P}
\frac{2\pi}{q}\int_{0}^{q} d p p\int_{|p-q|}^{|p+q|}d p' p'
\mathcal{I}^{o} &=&
\frac{7\pi^{3}|\Gamma(\mu)|^{4}q^{3}m^{2}}{3(2\pi)^{8}H^{2}}
\end{eqnarray}
Substituting eqs.  \eqref{P} into  \eqref{for10}, we have the
$\zeta$ correlation function due to a massive fermion loop as
\begin{eqnarray}\label{ZetaIn}
\int d^{3}x e^{i \textbf{q}\cdot
(\textbf{x}-\textbf{x}')}\Big\langle
\zeta(\textbf{x},t)\zeta(\textbf{x}',t)\Big\rangle_{loop} =
-\frac{28\pi^{3}|\Gamma(\mu)|^{4}(8\pi
G)^{2}m^{2}H^{2}}{3(2\pi)^{5}q^{3}}
\end{eqnarray}
Using $|\Gamma(\mu)|^{2} = |\Gamma(\frac{1}{2}\pm i r)|^{2} =
\frac{\pi}{\cosh \pi r}$, we have
\begin{eqnarray}\label{ZetaInIn}
\int d^{3}x e^{i \textbf{q}\cdot
(\textbf{x}-\textbf{x}')}\Big\langle
\zeta(\textbf{x},t)\zeta(\textbf{x}',t)\Big\rangle_{loop} =
-\frac{7(8\pi G)^{2}m^{2}H^{2}}{24 q^{3}\cosh^{2}\frac{\pi m}{H}}
\end{eqnarray}
which goes as \textit{$q^{-3}$ } at low momentum.\\

It should be noted that the power spectrum will \textit{not} be
zero when we take $m=0$.  Eq. \eqref{ZetaInIn} for massive
fermions is the result when the fermion pairs
$\bar{\psi}_{p'}\psi_{p}$ exit the horizon \textit{before or at
the same time} as $\zeta_{q}$ exits the horizon ($p \leq q$).  We
keep only the most dominant mode solutions for massive fermion
after horizon exit.  For the massless fermion case, the solution
is simple enough so that we can do the integration exactly without
any approximation.  The integrand contributed by the massive
fermion $\mathcal{F}^{o} \equiv
a_{1}^{3}a_{2}^{3}\mathcal{M}_{\psi}$ becomes \textit{frozen}
after horizon exit.  The negative power of $(-\tau)$ that arises
in the time integrals $\int\int dt_{2} dt_{1}$ of the massive
fermion, but not the massless fermion, is more important than the
exponential function when $\tau \rightarrow 0$.  Therefore,
\textit{massive fermion loops can contribute the $(\log a)^{2}$
factor} because the interaction goes as $a^{0}$.  In comparison,
$\log a$ does not arise in the massless fermion case because the
interaction goes as $a^{-1}$
rather than $a^{0}$. \\

As mentioned in section II, a fermion mass could arise from the
non-zero vacuum expectation value of an inflaton field in a flat
potential.  Therefore, the effective fermion mass during inflation
could be as large as $M_{Pl}$.  We know that in order to generate
all matter observed today, the inflaton $\varphi$ must couple to
matter such as fermions sometime during inflation.  Since we work
in the gauge where an inflaton does not fluctuate $\delta\varphi
=0$, the Yukawa coupling that can arise for the general inflaton
potential does not change the result in eq. \eqref{ZetaInIn} but
only shifts the fermion mass to be
\begin{equation}\label{m}
m \rightarrow m + \bar{\varphi}(t_{q})
\end{equation}
However, during inflation, the fermion mass could be large because
the non-zero expectation value of the scalar field
$\bar{\varphi}(t)$ could be large.  If the unperturbed inflaton
amplitude at the time of horizon exit is as large as $M_{Pl}$, we
have $m \simeq M_{Pl} \equiv \frac{1}{\sqrt{8\pi G}}$.  Therefore,
the power spectrum due to massive fermion loops is
\begin{eqnarray}\label{ZetaInInslowroll}
\int d^{3}x e^{i \textbf{q}\cdot
(\textbf{x}-\textbf{x}')}\Big\langle
\zeta(\textbf{x},t)\zeta(\textbf{x}',t)\Big\rangle_{loop, m=
M_{Pl}} = -\frac{7(8\pi G)H^{2}}{24 q^{3}\cosh^{2}\frac{\pi
M_{Pl}}{H}}
\end{eqnarray}
We see that even when we include the large $\bar{\varphi}\sim
M_{Pl}$ coupling that seems to give the quantum effect that does
not get suppressed by the factor of $G$, the result is suppressed
by the factor $\cosh^{2}\frac{M_{Pl}\pi}{H}$ instead.  This
happens because the inflaton fluctuates around a non-zero
background, implying that the massive fermion propagators are of
the order of $M_{Pl}$.  The factor $\cosh\frac{\pi m}{H}$ arises
when we fix
the mass dependent coefficients of the mode solution at late time.\\

It should be mentioned here that the large mass term \textit{does
not }get suppressed to the quantity like
$a^{3}\langle\bar{\psi}\psi\rangle \rightarrow \tanh \frac{\pi
m}{H}$.  Terms like $\bar{u}u$ or $\bar{v}v$ approach constants
with the mass dependent constant coefficient going as $\tanh
\frac{\pi m}{H}$, which does not have large mass suppression.
However, to close the fermion loop, the tree
$\langle\bar{\psi}\psi\rangle$ is not the only quantity we need to
calculate.  We also must calculate
$\langle\bar{\psi}_{1}\psi_{1}\bar{\psi}_{2}\psi_{2}\rangle$,
which is the trace over the multiplied matrices such as
$\sum_{r,s}\bar{v}_{p',r}(t_{1})u_{p,s}(t_{1})\bar{u}_{p,s}(t_{2})v_{p',r}(t_{2})$.
Therefore, the bilinear $\bar{v}_{1}u_{1}$ gives a constant mass
dependent coefficient of $\frac{1}{\cosh \frac{m\pi}{H}}$,
resulting in a small one-loop result even when the fermion couples
to the inflaton in the order of $M_{Pl}$.
\\

Apart from ultraviolet divergences, no infrared divergence can
arise due to the late time behavior.  The reason for infrared
safety comes from the fact that the function $\mathcal{F}^{o}$
approaches a constant at low momentum.  This is similar to viewing
$\sigma$ and $\bar{\sigma}$ in eq. \eqref{massfresult} as scalars
which approach constants after horizon exit.  Provided that the
integral over time is infrared safe, the integral over time only
comes from the $\zeta$ correlator, whereas the fermionic part only
contributes an $a^{-3}$ factor that always cancels with the factor
$\sqrt{-\bar{g}}$ in each interaction Hamiltonian.  After all
integrations, the power spectrum gives a $q^{-3}$ momentum
dependence.  The result of the massive fermion case is valid at
low momentum modes only where we need to cut off the momentum
integral $p$ to some value i.e., $\Lambda q$, so that the
approximation of small $p,p'$ in the Hankel function (eq.
\eqref{approx}) is still valid.  This case means that, if
$\Lambda$ is of order $1$, the fermion momentum modes $p,p'$ exit
the horizon \textit{before} the $\zeta$ momentum mode $q$ crosses
the horizon ($p,p' \leq q = a(t_{q})H(t_{q})$).  For higher
momentum modes $p$, the fermion behaves like it is massless and is
always suppressed by the factor $G$ and a negative power of the
Robertson Walker $a$ as shown in massless fermion section.
\\

To investigate whether the quantum effect is truly small, more
careful consideration is needed for the mass effect.  The reason
is that various mass dependent coefficients can arise when
matching the general solution with that inside the horizon for the
general graphs.  However, a fermion has two components that are
needed to form a pair with its conjugate.  Bilinears like
$\bar{u}u$ or $\bar{v}v$, but \textit{not} $\bar{v}u$, contribute
a constant factors like $|\Gamma(\mu)|^{2}(e^{\frac{\pi m}{H}}-
e^{-\frac{\pi m}{H}})$ which are $\tanh\frac{\pi m}{H}\rightarrow
1$ in the large mass limit.  Bilinears like $u^{\dag}u$ or
$v^{\dag}v$ contribute a constant factor like
$\frac{\cosh\frac{m\pi}{H}}{\cosh\frac{m\pi}{H}} = 1$, which is
mass independent.  Therefore, by considering this alone, a fermion
has no exponential suppression and seems to give a large quantum
effect if a vertex is as large as $M_{Pl}$.  However, as shown in
the detailed calculation here, this is not possible for the loop
graph that has two external legs with two trilinear vertices
because it requires a bilinear like $\bar{v}u$ instead.  Bilinear
terms like
 $\bar{\psi}\gamma^{i}\psi$ get suppressed at late time because $\gamma^{i}$
 can only have a contraction with $\frac{q_{i}}{a}$.  Therefore, the result is suppressed by
 an additional negative power of $a$ and its low momentum outside the horizon.
 Interaction terms like
 $\bar{\psi}\gamma^{0}\dot{\psi}$ contribute both $\frac{q_{i}}{a}\bar{\psi}\gamma^{i}\psi$
 and $m \bar{\psi}\psi$ factors via Dirac equation and its conjugate.  Therefore,
 the maximum result of $\bar{\psi}\gamma^{0}\dot{\psi}$ cannot exceed
 the result of $\bar{\psi}\psi$.  The other powers of bilinear
terms like $(\bar{\psi}\psi)^{n}$ for $n> 1$ cannot couple to the
mass dimension in the order of $M_{Pl}$ and can only give higher
fermion loops(by dimension counting in the action). Hence we
expect the $(\bar{\psi}\psi)^{n> 1}$ interaction type to be
suppressed by a negative power of $a(t)$.
\vspace{12pt}
\begin{center}
{\bf VI. VECTOR LOOP, INFLATON, AND GRAVITY}
\end{center}
\nopagebreak
Ford considered a classical vector field driving
inflation [12].  In this section, we consider a \textit{quantized}
vector field
  that affects the quantity $\zeta$ and its correlation function
  through the interaction with
  gravitational fluctuations.
 The action is
\begin{eqnarray}
\mathcal{L} &=&
\mathcal{L}_{g}+\mathcal{L}_{\varphi}+\mathcal{L}_{V} \nonumber \\
&=& -\frac{\sqrt{-g}}{2}\Big[\frac{1}{8\pi G} R
+g^{\mu\nu}\partial_{\mu}\varphi\partial_{\nu}\varphi +
2V(\varphi) \nonumber \\&&
+\frac{1}{2}g^{\mu\alpha}g^{\nu\beta}F_{\mu\nu}F_{\alpha\beta} +
m^{2}g^{\mu\nu}A_{\mu}A_{\nu}\Big]
\end{eqnarray}
To determine the vector field propagator, we need to solve the
interaction free field equation in an inflating universe.  It is
\begin{eqnarray}\label{FEvec}
\partial_{\mu}\Big(a^{3}(t)F^{\mu\nu}\Big)-a^{3}(t)m^{2}A^{\nu}= 0
\end{eqnarray}
For $\nu = 0$, this gives
\begin{eqnarray}\label{A0}
A_{0} = -\frac{iq_{i}\dot{A}_{i}}{q^{2}+(m a)^{2}}
\end{eqnarray}
For $\nu = j$, this gives
\begin{eqnarray}\label{FEvec1}
\ddot{A}_{j} + H \dot{A}_{j} + \frac{q^{2}}{a^{2}}A_{j}+
m^{2}A_{j} = \frac{q_{j}q_{i}A_{i}}{a^{2}}+ i q_{j}(\dot{A}_{0}+ H
A_{0})
\end{eqnarray}
To eliminate the auxiliary field $A_{0}$, we apply
$\partial_{\nu}$ to eq. \eqref{FEvec}.  We have
\begin{eqnarray}
\partial_{\nu}(a^{3}A^{\nu}) = 0
\end{eqnarray}
or
\begin{eqnarray}\label{FEvec2}
\dot{A}_{0}+ 3H A_{0} -\frac{i q_{i}}{a^{2}}A_{i} = 0
\end{eqnarray}
Substituting  \eqref{FEvec2} in \eqref{FEvec1}, we have the
dynamical field equation of $A_{j}$ in an expanding universe as
\begin{eqnarray}
\ddot{A}_{j} + H\Big(1 + \frac{2q_{i}q_{j}}{q^{2}+ (m
a)^{2}}\Big)\dot{A}_{j} + \Big(\frac{q^{2}}{a^{2}}+
m^{2}\Big)A_{j} = 0
\end{eqnarray}
For the transverse direction $q_{i}A_{i} = 0$, we have
\begin{eqnarray}
\ddot{A}_{j} + H\dot{A}_{j} + \Big(\frac{q^{2}}{a^{2}}+
m^{2}\Big)A_{j} = 0
\end{eqnarray}
where this is valid for photons ($m = 0$) and massive vector
bosons
in the transverse direction ($\lambda = 1, 2$).\\

For the parallel direction ($\lambda = 3, m \neq 0$ only), we have
\begin{eqnarray}\label{FEvecpara}
\ddot{A}_{j} + \Big(1+ \frac{2q^{2}}{q^{2}+(m
a)^{2}}\Big)\dot{A}_{j} + \Big(\frac{q^{2}}{a^{2}}+
m^{2}\Big)A_{j} = 0
\end{eqnarray}
It is impossible to solve equation \eqref{FEvecpara} exactly [13].
However, at late time during inflation, $a(t)$ grows more or less
exponentially.  Therefore, the second term proportional to
$\dot{A}_{j}$ in the equation above may be negligible.  Hence, the
vector field can be written as
\begin{eqnarray}
A_{i}(\textbf{x}, t) &=& \int d^{3}q \sum_{\lambda }
\Big[e^{i\textbf{q}\cdot\textbf{x}}e_{i}(\hat{q},\lambda)\alpha(\textbf{q},\lambda)\mathcal{A}_{q}(t)
\nonumber \\&&+ e^{-i\textbf{q}\cdot\textbf{x}}e^{*}_{i}
(\hat{q},\lambda)\alpha^{*}(\textbf{q},\lambda)\mathcal{A}^{*}_{q}(t)\Big]
\end{eqnarray}
where
\begin{eqnarray}
\sum_{\lambda, \lambda' =
1}^{2}e_{i}^{*}(\hat{q},\lambda)e_{j}(\hat{q},\lambda') =
\delta_{ij}-\hat{q}_{i}\hat{q}_{j}
\end{eqnarray}
for photons $m = 0$ and
\begin{eqnarray}
\sum_{\lambda, \lambda' =
1}^{3}e_{i}^{*}(\hat{q},\lambda)e_{j}(\hat{q},\lambda')
\rightarrow \delta_{ij}
\end{eqnarray}
for massive vector bosons during late time inflation.  Therefore,
$\mathcal{A}_{q}(t)$ is the solution that satisfies
\begin{equation}
\frac{d}{dt}\Big(a(t)\frac{d}{dt}\mathcal{A}_{q}(t)\Big) +
\frac{q^{2}}{a(t)}\mathcal{A}_{q}(t) + m^{2} a \mathcal{A}_{q}(t)
= 0
\end{equation}
To solve the equation above at a general momentum $q$, we can work
in the conformal time $\tau$.  Hence the massive gauge field
equation in an inflating universe is
\begin{equation}\label{gaugeeq}
\frac{d^{2}\mathcal{A}_{q}}{d\tau^{2}} + \Big( q^{2} +
\frac{r^{2}}{\tau^{2}}\Big)\mathcal{A}_{q} = 0
\end{equation}
where $r \equiv \frac{m}{H}$.\\

We see from eq. \eqref{gaugeeq} that in the limit of $m = 0$, the
solution of a massless vector field is a plane wave.  This
solution is the same as those of the conformal scalar and massless
fermion.  The positive mode solution for a massless vector field
at general wavelength is
\begin{equation}
\mathcal{A}_{q}(t) = \frac{1}{(2\pi)^{\frac{3}{2}}\sqrt{2q}}e^{-i
q\tau}, m = 0
\end{equation}
For a massive vector field $m\neq 0$, the field equation
\eqref{gaugeeq} has the Bessel's equation type [11]
\begin{equation}
u''_{q} + \Big(q^{2}- \frac{4\nu^{2}-1}{4\tau^{2}})u_{q} = 0
\end{equation}
Therefore, the general solution of a massive vector field is
\begin{equation}
\mathcal{A}_{q}(\tau) =
\mathcal{E}_{q}\sqrt{-\tau}H^{(1)}_{\nu}(-q\tau)+
\mathcal{F}_{q}\sqrt{-\tau}H^{(2)}_{\nu}(-q\tau), m\neq 0
\end{equation}
where
\begin{equation}
\nu = \sqrt{\frac{1}{4}- r^{2}}
\end{equation}
Since we want the solution to match the positive solution at deep
inside the horizon $e^{- i \omega \tau}$, only $H^{(1)}_{\nu}(x)$
but not $H^{(2)}_{\nu}(x)$ gives an $e^{-i\omega \tau}$ factor in
the large $|x|$ limit.  Hence, $\mathcal{F}_{q}= 0$ and
\begin{eqnarray}\label{modesol}
\mathcal{A}_{q}(t) &=& \mathcal{E}_{q}\sqrt{-\tau}
H^{(1)}_{\nu}(-q\tau)
\end{eqnarray}
A normalized constant $\mathcal{E}_{q}$ is chosen to match with
the solution at deep inside the horizon.  Inside the horizon, the
positive frequency solution is the same as that in flat space,
which is,
\begin{eqnarray}\label{inside}
\mathcal{A}_{q}(t) &\rightarrow&
\frac{1}{(2\pi)^{\frac{3}{2}}\sqrt{2\omega_{q}}}
\exp\Big(-i\int^{\tau}_{-\infty}\omega_{q}(\tau')d\tau'\Big)
\end{eqnarray}
where $\omega_{q}(\tau) \equiv \sqrt{ q^{2}+ (m a)^{2}}$.  With
the property of Hankel's function in the asymptotic limit,
$|x|\rightarrow \infty$
\begin{equation}\label{largexlimit}
H_{\nu}^{(1)}(x)\rightarrow \sqrt{\frac{2}{\pi
x}}\exp\Big(i(x-\frac{\pi \nu}{2}-\frac{\pi}{4})\Big)
\end{equation}
Since we now allow the existence of a mass term which can be
either large or small, the normalized constants $\mathcal{E}_{q}$
can be
 a function of mass and this may affect the result of the correlation
function.\\

During inflation, the mass of the matter could be large due to the
interaction of matter with the inflaton $\bar{\varphi}$.  As
mentioned earlier, the slow roll condition of some inflationary
theories requires $m = \bar{\varphi}\simeq M_{Pl}$.  This can make
the mass term as large as $M_{Pl}$ and may affect the final result
of the correlation function.  To determine the mass dependent
coefficient $\mathcal{E}_{q}$, we match the solution with that
inside the horizon.  From eqs. \eqref{modesol},  \eqref{inside},
and
 \eqref{largexlimit}, we have the mass dependent coefficient
$\mathcal{E}_{q}$ as
\begin{eqnarray}\label{coeff}
\mathcal{E}_{q}(m) &=&
\frac{\sqrt{\pi}}{2(2\pi)^{\frac{3}{2}}}e^{\frac{i\pi}{4}(1+
2\nu)}
\end{eqnarray}
From eqs. \eqref{modesol} and \eqref{coeff}, we therefore have the
massive mode solution of the gauge field that we will use for the
propagator as
\begin{eqnarray}\label{modesolA}
\mathcal{A}_{q}(t) &=&
\frac{\sqrt{\pi}}{2(2\pi)^{\frac{3}{2}}}e^{\frac{i\pi}{4}(1+
2\nu)}\sqrt{-\tau} H^{(1)}_{\nu}(-q\tau)
\end{eqnarray}
where $\nu = \sqrt{\frac{1}{2}-r^{2}}$. \\

To calculate what the interaction vertices are, we need to expand
beyond quadratic order in fluctuations.  As derived in the
appendix, the cubic order is
\begin{eqnarray}\label{vecint}
H_{\zeta A A}(t) = -\int d^{3}x \epsilon H a^{5}\Big(
\frac{1}{a^{2}}\dot{A}_{i}^{2} +
\frac{1}{2a^{4}}(\partial_{i}A_{j}-\partial_{j}A_{i})^{2}\Big)\nabla^{-2}\dot{\zeta}
\end{eqnarray}
where we use the energy momentum tensor in [7] and choose gauge
$A_{0}=0$ for $m=0$.  For $m\neq0$, we can solve the constraint
equation of $A_{0}$ and plug it back into the action.  As seen
from eq. \eqref{A0}, $A_{0}$ is decaying as
$\frac{q_{i}\dot{A}_{i}}{(am)^{2}}$ after horizon exit. Therefore,
we can approximate the massive vertices as that of the massless
case.  It is only the propagators that will be different from
massless case.  To continue, we use the general formula in
eq.\eqref{genformula} with the replacement
\begin{eqnarray}
\Psi^{*}\Psi &\rightarrow& \frac{1}{a^{2}}\dot{A}_{i}^{2} +
\frac{1}{a^{4}}\Big((\partial_{i}A_{j})^{2}-\partial_{i}A_{j}\partial_{j}A_{i}\Big)
\\
\zeta_{q}(t_{1,2})&\rightarrow& -\dot{\zeta}_{q}(t_{1,2})/q^{2}\\
V(t) &=& -\epsilon H a^{5}(t)
\end{eqnarray}
Since the purely electric term (through $(\dot{A}_{i})^{2}$),
purely magnetic term (through
$(\partial_{i}A_{j})^{2}-(\partial_{i}A_{j})(\partial_{j}A_{i})$),
and two cross terms arise when we evaluate the commutator
$[H_{1},[H_{2},Q]]$, eq. \eqref{matterA} becomes
\begin{eqnarray}
\mathcal{M}_{\mathcal{A}}
&=&\frac{\mathcal{P}_{1}}{a_{1}^{2}a_{2}^{2}}
\dot{\mathcal{A}}_{p}(t_{1})\dot{\mathcal{A}}_{p'}(t_{1})
\dot{\mathcal{A}}^{*}_{p}(t_{2})\dot{\mathcal{A}}^{*}_{p'}(t_{2})+
\nonumber
\\&&\frac{\mathcal{P}_{2}}{a_{1}^{4}a_{2}^{4}}\mathcal{A}_{p}(t_{1})\mathcal{A}_{p'}(t_{1})
\mathcal{A}^{*}_{p}(t_{2})\mathcal{A}^{*}_{p'}(t_{2})+ \nonumber
\\
&&\frac{\mathcal{P}_{3}}{a_{1}^{4}a_{2}^{2}}\mathcal{A}_{p}(t_{1})\mathcal{A}_{p'}(t_{1})
\dot{\mathcal{A}}^{*}_{p}(t_{2})\dot{\mathcal{A}}^{*}_{p'}(t_{2})
+\nonumber
\\&&\frac{\mathcal{P}_{4}}{a_{2}^{3}a_{4}}\dot{\mathcal{A}}_{p}(t_{1})\dot{\mathcal{A}}_{p'}(t_{1})
\mathcal{A}^{*}_{p}(t_{2})\mathcal{A}^{*}_{p'}(t_{2})
\end{eqnarray}
and $\mathcal{\zeta}$ remains the same as in eq. \eqref{zeta}
because we are calculating the same correlation function $\langle
\zeta\zeta \rangle$ with various kinds of matter loops.
\\

We now need to calculate the polarization factor $\mathcal{P}_{i}$
for $i = 1...4$. \\ The $\mathcal{P}_{1}$ factor comes from the
purely electric field term which is
\begin{eqnarray}
\mathcal{P}_{1} &=&
\sum_{\lambda,\lambda'=1}^{2}(e_{i,p,\lambda}e^{*}_{j,p,\lambda})
(e_{i,p',\lambda'}e^{*}_{j,p',\lambda'})\nonumber \\
&=& 1+ (\hat{p}\cdot \hat{p}')^{2}
\end{eqnarray}
The $\mathcal{P}_{2}$ factor comes from the purely magnetic field
term which is
\begin{eqnarray}
\mathcal{P}_{2} &=&
\sum_{\lambda,\lambda'=1}^{2}|(\textbf{p}\cdot\textbf{p}')(\hat{e}_{p,\lambda}\cdot
\hat{e}_{p',\lambda'})-(\textbf{p}\cdot
\hat{e}_{p',\lambda'})(\textbf{p}'\cdot
\hat{e}_{p,\lambda})|^{2}\nonumber \\
&=& p^{2}p'^{2}(1+(\hat{p}\cdot \hat{p}')^{2})
\end{eqnarray}
The $\mathcal{P}_{3,4}$ factors come from the cross terms which
are
\begin{eqnarray}
\mathcal{P}_{3} = \mathcal{P}_{4} = - 2 \textbf{p}\cdot
\textbf{p}'
\end{eqnarray}
Substitute these into eq. \eqref{genformula}, we get
\begin{eqnarray}\label{gaugeresult}
\int d^{3}x e^{i \textbf{q}\cdot
(\textbf{x}-\textbf{x'})}\Big\langle
\zeta(\textbf{x},t)\zeta(\textbf{x}',t)\Big\rangle_{loop} =
-8(2\pi)^{9}\int
d^{3}pd^{3}p' \delta^{3}(\textbf{q}+\textbf{p}+\textbf{p}') \nonumber \\
\times \int_{-\infty}^{t}dt_{2}\epsilon_{2}H_{2}a_{2}^{5}
\int_{-\infty}^{t_{2}}dt_{1}\epsilon_{1}H_{1}a_{1}^{5}Re\Big(\mathcal{Z}\mathcal{M}_{\mathcal{A}}
\Big)
\end{eqnarray}
where
\begin{eqnarray}\label{G}
\mathcal{M}_{A} &=&\Big(1+ (\hat{p}\cdot
\hat{p}')^{2}\Big)a_{1}^{-2}a_{2}^{-2}\dot{\mathcal{A}}_{p}(t_{1})\dot{\mathcal{A}}_{p'}(t_{1})
\dot{\mathcal{A}}^{*}_{p}(t_{2})\dot{\mathcal{A}}^{*}_{p'}(t_{2})
\nonumber \\&&+ p^{2}p'^{2}\Big(1+
(\hat{p}\cdot\hat{p}')^{2}\Big)a_{1}^{-4}a_{2}^{-4}\mathcal{A}_{p}(t_{1})\mathcal{A}_{p'}(t_{1})
\mathcal{A}^{*}_{p}(t_{2})\mathcal{A}^{*}_{p'}(t_{2}) \nonumber \\
&&-2
pp'(\hat{p}\cdot\hat{p}')a_{1}^{-4}a_{2}^{-2}\mathcal{A}_{p}(t_{1})\mathcal{A}_{p'}(t_{1})
\dot{\mathcal{A}}^{*}_{p}(t_{2})\dot{\mathcal{A}}^{*}_{p'}(t_{2})
\nonumber \\&&- 2
pp'(\hat{p}\cdot\hat{p}')a_{1}^{-2}a_{2}^{-4}\dot{\mathcal{A}}_{p}(t_{1})\dot{\mathcal{A}}_{p'}(t_{1})
\mathcal{A}^{*}_{p}(t_{2})\mathcal{A}^{*}_{p'}(t_{2})
\end{eqnarray}
Since the solution of the massless vector field is just a plane
wave,
\begin{equation}
\dot{\mathcal{A}}_{q}(t) = -\frac{i q}{a(t)}\mathcal{A}_{q}(t)
\end{equation}
eq. \eqref{G} can be simplified further as
\begin{eqnarray}\label{GG}
\mathcal{M}_{\mathcal{A}} &=&  \frac{2
p^{2}p'^{2}}{a_{1}^{4}a_{2}^{4}}\Big(1+
(\hat{p}\cdot\hat{p}')\Big)^{2}\mathcal{A}_{p}(t_{1})\mathcal{A}_{p'}(t_{1})
\mathcal{A}^{*}_{p}(t_{2})\mathcal{A}^{*}_{p'}(t_{2})\nonumber \\
&=&  \frac{ p^{2}p'^{2}}{2(2\pi)^{6}a_{1}^{4}a_{2}^{4}p p'}\Big(1+
(\hat{p}\cdot\hat{p}')\Big)^{2}e^{-i(p+p')(\tau_{1}-\tau_{2})}
\end{eqnarray}
Substituting the $\mathcal{Z}$ part in eq. \eqref{zetapart} and
gauge field part in eq. \eqref{GG} into eq. \eqref{gaugeresult},
we have the correlation function due to massless vector fields:
\begin{eqnarray}\label{gaugeresult1}
\int d^{3}x e^{i \textbf{q}\cdot
(\textbf{x}-\textbf{x'})}\Big\langle
\zeta(\textbf{x},t)\zeta(\textbf{x}',t)\Big\rangle_{loop} =
-16(2\pi)^{9}\epsilon^{2}|\zeta^{o}_{q}|^{4}\int
d^{3}pd^{3}p'\nonumber \\\times
\delta^{3}(\textbf{q}+\textbf{p}+\textbf{p}')|\mathcal{A}^{o}_{p}|^{2}
|\mathcal{A}^{o}_{p'}|^{2}p^{2}p'^{2}\Big(1+
(\hat{p}\cdot\hat{p}')\Big)^{2}\mathcal{T}
\end{eqnarray}
where
\begin{eqnarray}\label{timtint}
\mathcal{T}&=& Re\int_{-\infty}^{0}d\tau_{2}
(e^{-iq\tau_{2}}-e^{iq\tau_{2}})e^{i(p+p')\tau_{2}}\int_{-\infty}^{\tau_{2}}d\tau_{1}
e^{-i(q+p+p')\tau_{1}} \nonumber \\
&=& -\frac{1}{2q(q+p+p')}
\end{eqnarray}
and the constant coefficients after horizon exit are
\begin{eqnarray}\label{coeffA}
|\mathcal{A}^{o}_{q}|^{2} = \frac{1}{2 (2\pi)^{3}q}
\end{eqnarray}
Substituting eqs. \eqref{timtint} and  \eqref{coeffA} into
\eqref{gaugeresult1}, we have the loop power spectrum due to a
massless vector field as
\begin{eqnarray}\label{gaugeresult2}
\int d^{3}x e^{i \textbf{q}\cdot
(\textbf{x}-\textbf{x'})}\Big\langle
\zeta(\textbf{x},t)\zeta(\textbf{x}',t)\Big\rangle_{loop} =
\frac{(8\pi G H^{2})^{2}}{2(2\pi)^{3}q^{7}}\int d^{3}pd^{3}p'
\nonumber \\\times\delta^{3}(\textbf{q}+\textbf{p}+\textbf{p}')
\frac{pp'}{q+p+p'}\Big(1 + (\hat{p}\cdot\hat{p}')\Big)^{2}
\end{eqnarray}
Notice that the first term is the same as that of a massless
minimal coupled scalar loop.  The $\hat{p}\cdot\hat{p}'$ terms
come from summation over polarization vectors.  We can follow the
same method of dimensional regularization shown in the earlier
section.  We have the finite part of correlation function due to
massless gauge field as
\begin{eqnarray}\label{gaugedimpresult}
\int d^{3}x e^{i \textbf{q}\cdot
(\textbf{x}-\textbf{x}')}\Big\langle
\zeta(\textbf{x},t)\zeta(\textbf{x}',t)\Big\rangle_{loop} =
-\frac{ 44\pi(8\pi G H^{2}(t_{q}))^{2}}{30(2\pi)^{3}q^{3}}
\Big(\ln q + C\Big)
\end{eqnarray}
Notice that the result is smaller than the classical result by a
factor of $G H^{2}$ in order of magnitude.  The numerical
coefficient is slightly more than that of a scalar loop in [2]
because there are additional polarization factors.
\vspace{12pt}
\begin{center}
{\bf VII.  MASSIVE VECTOR BOSON}
\end{center}
\nopagebreak
We consider the late time mode solution of the
massive vector field because the mode solution for the propagator
is no longer a simple plane wave as in the massless case.  For
$|x|= |-p\tau| \rightarrow 0$, we have
\begin{equation}
J_{\nu}(x) \rightarrow \frac{x^{\nu}}{2^{\nu}\Gamma(\nu+1)}\Big(1+
\mathcal{O}(x^{2})\Big)
\end{equation}
By definition of Hankel's function,
\begin{equation}
H_{\nu}^{(1)}(x) = \frac{1}{i \sin \nu \pi}\Big(
J_{-\nu}(x)-e^{-\nu\pi i}J_{\nu}(x)\Big)
\end{equation}
Hence, the late time behavior of mode solution approaches
\begin{equation}\label{Hnu}
H_{\nu}^{(1)}(x) \rightarrow -\frac{i}{\sin \nu \pi}\Big(
\frac{x^{-\nu}}{2^{-\nu}\Gamma(1-\nu)}- \frac{e^{-\nu\pi
i}x^{\nu}}{2^{\nu}\Gamma(\nu+1)}\Big)\Big(1+
\mathcal{O}(x)^{2}\Big)
\end{equation}
The exact solution in eq. \eqref{modesolA} approaches
\begin{eqnarray}\label{Asol}
\mathcal{A}_{q}(t) = \mathcal{C}_{q}a^{\lambda_{+}}+
\mathcal{D}_{q}a^{\lambda_{-}}
\end{eqnarray}
at late time where
\begin{eqnarray}
\lambda_{\pm} = -\frac{1}{2}\pm \nu = -\frac{1}{2}\pm
\sqrt{\frac{1}{4}-\frac{m^{2}}{H^{2}}}
\end{eqnarray}
and
\begin{equation}\label{masscoeff}
\mathcal{C}_{q} = -\frac{i\sqrt{\pi}}{2(2\pi)^{\frac{3}{2}}\sqrt{
H}\sin\nu\pi }\frac{e^{\frac{i\pi}{4}(1+
2\nu)}}{\Gamma(1-\nu)}\Big(\frac{2H}{q}\Big)^{\nu}
\end{equation}
\begin{equation}\label{masscoeff1}
\mathcal{D}_{q}
=\frac{i\sqrt{\pi}}{2(2\pi)^{\frac{3}{2}}\sqrt{H}\sin\nu\pi}
\frac{e^{\frac{i\pi}{4}(1-
2\nu)}}{\Gamma(1+\nu)}\Big(\frac{2H}{q}\Big)^{-\nu}
\end{equation}
The time derivative of the solution contributes the same power of
$a$ as
\begin{eqnarray}\label{Adotsol}
\dot{\mathcal{A}}_{q}(t) =
H\Big(\lambda_{+}\mathcal{C}_{q}a^{\lambda_{+}}+
\lambda_{-}\mathcal{D}_{q}a^{\lambda_{-}}\Big)
\end{eqnarray}
We see from the equation above that the time derivative of the
propagator in massive theories contributes an additional factor
\begin{equation}
\dot{A}_{q}\rightarrow \lambda_{\pm}f(t) \rightarrow
\Big(-\frac{1}{2}\pm
\sqrt{\frac{1}{2}-\frac{m^{2}}{H^{2}}}\Big)f(t)
\end{equation}
where $f(t)$ has the same power of $t$ as in $\mathcal{A}_{q}$.
Therefore, there will be a term like
\begin{equation}
\dot{\mathcal{A}}_{p}(t_{1})\dot{\mathcal{A}}_{p'}(t_{1})
\dot{\mathcal{A}}^{*}_{p}(t_{2})\dot{\mathcal{A}}^{*}_{p'}(t_{2})
\rightarrow \mathcal{O}(\lambda^{4}_{\pm})F[t_{1},t_{2}]
\end{equation}
in the loop.  Since at large mass limit,
\begin{equation}\label{lam}
\lambda_{\pm}^{4}\rightarrow \frac{M_{Pl}^{4}}{H^{4}}
\end{equation}
the time derivative propagators give an additional factor of
$(\lambda_{\pm})^{4}$ for four fields.  Because of eq.
 \eqref{lam}, the loop spectrum does not seem to get suppressed by
an additional factor of $G$, but may get suppressed by the
constant coefficient at large mass in eqs.  \eqref{masscoeff} and
 \eqref{masscoeff1} or the results of the loop integrals.  We therefore investigate
 whether there is true suppression or not.\\

To analyze in more detail, we start from the interaction in eq.
\eqref{vecint}.  Therefore, the massive vector field loop
 contributes
\begin{eqnarray}\label{Mgauge1}
\mathcal{M}_{\mathcal{A}, m\neq 0} &=& \Pi^{*}(t_{2})\Pi(t_{1})
\end{eqnarray}
where
\begin{eqnarray}
\Pi(t) &=&
 \frac{1}{a^{2}(t)}\dot{A}_{i, p, \lambda}(t)\dot{A}_{i, p', \lambda'}(t)
 +\frac{1}{a^{4}(t)}\Big(p_{i}p'_{j}A_{j, p, \lambda}(t)A_{i, p', \lambda'}(t)
 \nonumber \\&&-
 p_{i}p'_{i} A_{j, p, \lambda}(t)A_{j, p', \lambda'}(t)\Big)
\end{eqnarray}
The exact solution of a massive vector field involves Hankel's
functions which are rather complicated to integrate over time and
momentums.  However, we can get some ideas about what the momentum
dependence of the observable spectrum is by considering the long
wavelength mode solutions.  We see from eqs. \eqref{Asol} and
\eqref{Adotsol} that $\dot{\mathcal{A}_{q}}$ gives the same power
of $a$ as $\mathcal{A}_{q}$ at late time.  Therefore, we can keep
the most leading order term as the universe rapidly expands
\begin{eqnarray}\label{Pi212}
\mathcal{M}_{A} &\rightarrow& \frac{3}
{a_{2}^{2}a_{1}^{2}}\dot{\mathcal{A}_{p}}^{*}(t_{2})
\dot{\mathcal{A}_{p'}}^{*}(t_{2})\dot{\mathcal{A}_{p}}(t_{1})
\dot{\mathcal{A}_{p'}}(t_{1})
\end{eqnarray}
This means that the (massive) electric-like term is more
dominating than the magnetic-like term after horizon exit.  This
result is different from the result of massless vector fields in
which all electric and magnetic terms are equally important.
Substituting eq. \eqref{Pi212} into eq. \eqref{genformula}, we
have
\begin{eqnarray}\label{gaugeresultmass}
\int d^{3}x e^{i \textbf{q}\cdot
(\textbf{x}-\textbf{x}')}\Big\langle
\zeta(\textbf{x},t)\zeta(\textbf{x}',t)\Big\rangle_{loop}
\rightarrow -24(2\pi)^{9}\int
d^{3}pd^{3}p' \delta^{3}(\textbf{q}+\textbf{p}+\textbf{p}') \nonumber \\
\int_{-\infty}^{t}dt_{2}\epsilon_{2}H_{2}a_{2}^{3}
\int_{-\infty}^{t_{2}}dt_{1}\epsilon_{1}H_{1}a_{1}^{3}
Re\Big(\mathcal{Z} \dot{\mathcal{A}_{p}}^{*}(t_{2})
\dot{\mathcal{A}_{p'}}^{*}(t_{2})\dot{\mathcal{A}_{p}}(t_{1})
\dot{\mathcal{A}_{p'}}(t_{1})\Big)
\end{eqnarray}
where $\mathcal{Z}$ is the contribution from the $\zeta$ part
which is still the same as that in eq. \eqref{zetapart}.  Hence,
\begin{eqnarray}\label{gaugeresultmass1}
\int d^{3}x e^{i \textbf{q}\cdot
(\textbf{x}-\textbf{x}')}\Big\langle
\zeta(\textbf{x},t)\zeta(\textbf{x}',t)\Big\rangle_{loop}
\rightarrow -24(2\pi)^{9}|\epsilon|^{2}|\zeta^{o}_{q}|^{4}\int
d^{3}pd^{3}p'
\delta^{3}(\textbf{q}+\textbf{p}+\textbf{p}')\mathcal{T}
\end{eqnarray}
where
\begin{eqnarray}\label{Ttotal}
\mathcal{T}&\equiv& Re\int_{-\infty}^{t}dt_{2}a_{2} ( e^{
-iq\tau_{2}}-e^{i q \tau_{2}}) \dot{\mathcal{A}_{p}}^{*}(t_{2})
\dot{\mathcal{A}_{p'}}^{*}(t_{2}) \nonumber
\\&&\times\int_{-\infty}^{t_{2}}dt_{1}a_{1} e^{-i
q\tau_{1}}\dot{\mathcal{A}_{p}}(t_{1})
\dot{\mathcal{A}_{p'}}(t_{1})
\end{eqnarray}
With the time derivative of the late time mode solution in eq.
\eqref{Adotsol}, we have
\begin{eqnarray}\label{Pilate}
\dot{\mathcal{A}_{p}}(t_{1})\dot{\mathcal{A}_{p'}}(t_{1}) &=&
H^{2}\Big[\lambda_{+}^{2}\mathcal{C}_{p}\mathcal{C}_{p'}a_{1}^{2\lambda_{+}}
+\lambda_{-}^{2}\mathcal{D}_{p}\mathcal{D}_{p'}a_{1}^{2\lambda_{-}}
\nonumber
\\&&+\lambda_{+}\lambda_{-}(\mathcal{C}_{p}\mathcal{D}_{p'}+
\mathcal{D}_{p}\mathcal{C}_{p'})a_{1}^{\lambda_{+}+\lambda_{-}}\Big]
\end{eqnarray}
Therefore, the $t_{1}$ integral is
\begin{eqnarray}\label{T1}
\int_{-\infty}^{t_{2}}dt_{1}a_{1} e^{-i
q\tau_{1}}\dot{\mathcal{A}_{p}}(t_{1})
\dot{\mathcal{A}_{p'}}(t_{1})= H^{2}
\int_{-\infty}^{t_{2}}dt_{1}e^{-i q\tau_{1}} \Big[
\lambda_{+}^{2}\mathcal{C}_{p}\mathcal{C}_{p'}a_{1}^{2\nu}
\nonumber
\\+\lambda_{-}^{2}\mathcal{D}_{p}\mathcal{D}_{p'}a_{1}^{-2\nu}
+\lambda_{+}\lambda_{-}(\mathcal{C}_{p}\mathcal{D}_{p'}+
\mathcal{D}_{p}\mathcal{C}_{p'}) \Big]\nonumber \\
\rightarrow H\Big[ \frac{\lambda_{+}^{2}}{2\nu } c_{+}a_{2}^{2\nu}
-\frac{ \lambda_{-}^{2}}{2\nu }c_{-}a_{2}^{-2\nu} -
\lambda_{+}\lambda_{-}c_{0}Ei(-iq\tau_{2})\Big]
\end{eqnarray}
for $2\nu\neq 0$ and
\begin{eqnarray}\label{c0pm}
c_{0} &=& \mathcal{C}_{p}\mathcal{D}_{p'}+
\mathcal{D}_{p}\mathcal{C}_{p'}\\
c_{+} &=& \mathcal{C}_{p}\mathcal{C}_{p'}\\
c_{-}&=& \mathcal{D}_{p}\mathcal{D}_{p'}
\end{eqnarray}
and $\lambda_{\pm} = -\frac{1}{2}\pm
(\nu=\sqrt{\frac{1}{4}-\frac{m^{2}}{H^{2}}})$ which can be either
real or complex or zero, depending on its mass when compared to
the expansion rate $H$.\\\\
\textbf{ Small Mass: $m < \frac{H}{2}$}\\\\
For $m < \frac{H}{2}$, $\nu$ is real.  Therefore,
\begin{equation}
-\frac{1}{2}<\lambda_{+}< 0, -1 <\lambda_{-}< -\frac{1}{2}
\end{equation}
and $|\mathcal{C}_{q}|^{2}$ and $|\mathcal{D}_{q}|^{2}$ are
\textit{$q$-dependent}, depending on its mass.  Hence,
\begin{eqnarray}\label{Pilatetime1}
\mathcal{T} &=& H^{3}Re\int_{-\infty}^{t}dt_{2} (-2i)\sin
q\tau_{2}\Big[ \lambda_{+}^{2} c_{+}^{*}a_{2}^{2\nu}
+\lambda_{-}^{2}c_{-}^{*}a_{2}^{-2\nu}
+\lambda_{+}\lambda_{-}c^{*}_{0} \Big] \nonumber \\&& \times \Big[
\frac{\lambda_{+}^{2}}{2\nu } c_{+}a_{2}^{2\nu}
-\frac{\lambda_{-}^{2}}{2\nu }c_{-}a_{2}^{-2\nu}
-\lambda_{+}\lambda_{-}c_{0}Ei(-iq\tau_{2}) \Big]
\end{eqnarray}
We can see that there is no contribution from the terms
proportional to $|c_{+}|^{2}$, and $|c_{-}|^{2}$ because they are
all real. With the factor $i$ in the integrand, the contribution
is purely imaginary.  Hence there is no contribution after taking
the real part.  Therefore, the terms that give non-zero result are
\begin{eqnarray}
\mathcal{T} &=& -\frac{H^{2}\lambda_{+}\lambda_{-}}{\nu}Re
\int_{-\infty}^{\tau}\frac{d\tau_{2}}{\tau_{2}}  i  \sin q\tau_{2} \nonumber\\
 &&\Big[
2\nu\Big(\frac{\lambda_{+}^{2}c^{*}_{+}c_{0}}{(-H
\tau_{2})^{2\nu}}+ \frac{\lambda_{-}^{2}c^{*}_{-}c_{0}}{(-H
\tau_{2})^{-2\nu}}-\lambda_{+}\lambda_{-}|c_{0}|^{2}\Big)
Ei(-iq\tau_{2}) \nonumber
\\&&
-2\lambda_{+}\lambda_{-}i Im (c^{*}_{-}c_{+}) -
\frac{\lambda_{+}^{2}c^{*}_{0}c_{+}}{(-H \tau_{2})^{2\nu}} +
\frac{\lambda_{-}^{2}c^{*}_{0}c_{-}}{(-H \tau_{2})^{-2\nu}}\Big]
\end{eqnarray}
The integral above is still too complicated.  However, when  $0
\ll 2\nu < 1 $, $\sin q\tau_{2} \rightarrow q\tau_{2} =
-\frac{q}{a_{2}H}$, and $Ei(-i x) \simeq \ln x $ when
$x\rightarrow 0$,  the result of the time integral is less than
$t^{2}$ or $(\ln a(t))^{2}$ at late time.  Therefore, the dominant
contribution comes from the first term of the integrand.  Hence,
\begin{eqnarray}
\mathcal{T} \rightarrow 2q H \lambda_{-}
\lambda_{+}^{3}Im(c^{*}_{+}c_{0})\frac{(1-\eta\ln
q\tau)}{a^{\eta}\eta^{2}}
\end{eqnarray}
for $0 \ll 2\nu < 1$.  Therefore, the correlation function due to
a massive vector field loop is
\begin{eqnarray}\label{veclessmass}
\int d^{3}x e^{i \textbf{q}\cdot
(\textbf{x}-\textbf{x}')}\Big\langle
\zeta(\textbf{x},t)\zeta(\textbf{x}',t)\Big\rangle \rightarrow
-\frac{12(2\pi)^{3}\lambda_{-} \lambda_{+}^{3}(8\pi G
H^{2})^{2}H}{q^{5}a^{\eta}} \nonumber \\\times\frac{(1-\eta\ln
q\tau)}{\eta^{2}}\int d^{3}pd^{3}p'
\delta^{3}(\textbf{q}+\textbf{p}+\textbf{p}')Im(c^{*}_{+}c_{0})
\end{eqnarray}
To calculate what $I m(c^{*}_{+}c_{0})$ is, we see from eqs.
\eqref{masscoeff} and \eqref{masscoeff1} that
$\mathcal{C}^{*}_{p}\mathcal{D}_{p}$ is $p$-independent.
Therefore,
\begin{eqnarray}
I m (c^{*}_{+}c_{0}) = \frac{(2H)^{2\nu}\Gamma^{2}(\nu)}{8\nu
(2\pi)^{7}H^{2}}\Big[\frac{1}{p^{2\nu}}+\frac{1}{p'^{2\nu}}\Big]
\end{eqnarray}
where we use $\Gamma(1-x)\Gamma(x) = \frac{\pi}{\sin \pi x}$ and
\begin{eqnarray}
 I m (\mathcal{C}^{*}_{p}\mathcal{D}_{p}) &=& \frac{1}{4(2\pi)^{3}H
 \nu}\\
 |\mathcal{C}_{p}|^{2} &=& \frac{\pi
 \Gamma^{2}(\nu)}{(2\pi)^{5}H}\Big(\frac{2H}{p}\Big)^{2\nu}
\end{eqnarray}
Hence,
\begin{eqnarray}\label{vecresultSM1}
&&\int d^{3}x e^{i \textbf{q}\cdot
(\textbf{x}-\textbf{x'})}\Big\langle
\zeta(\textbf{x},t)\zeta(\textbf{x}',t)\Big\rangle_{loop, m<
\frac{H}{2}} = \nonumber \\&&-\frac{3
(2)^{2\nu}\Gamma^{2}(\nu)\lambda_{-} \lambda_{+}^{3}(8\pi G
H^{2})^{2}}{2(2\pi)^{4}H^{1-2\nu}
q^{5}a^{1-2\nu}}\frac{(1+(2\nu-1)\ln q\tau)}{\nu(1-2\nu)^{2}}
\nonumber
\\&&
\times \int d^{3}pd^{3}p'
\delta^{3}(\textbf{q}+\textbf{p}+\textbf{p}')
\Big[\frac{1}{p^{2\nu}}+\frac{1}{p'^{2\nu}}\Big]
\end{eqnarray}
To determine the momentum dependence $q$ of the spectrum, we
integrate over internal momentum $p, p'$ circulated inside the
loop.  Following the similar way as in the massive fermion
section, we have the momentum dependent spectrum as
\begin{eqnarray}\label{vecresultSM5}
\int d^{3}x e^{i \textbf{q}\cdot
(\textbf{x}-\textbf{x}')}\Big\langle
\zeta(\textbf{x},t)\zeta(\textbf{x}',t)\Big\rangle= \nonumber
\\-\frac{24 \Gamma^{2}(\nu)\lambda_{-} \lambda_{+}^{3}(8\pi G
H^{2})^{2}}{(2\pi)^{3}(2 H a(t))^{\eta}}\frac{(1-\eta\ln
q\tau)}{(1-\eta)(2+\eta)\eta^{2} q^{3-\eta}}
\end{eqnarray}
where
\begin{eqnarray}
2\nu = \sqrt{1-\frac{4 m^{2}}{H^{2}}} \simeq
1-\frac{2m^{2}}{H^{2}}
\end{eqnarray}
or
\begin{eqnarray}
\eta \equiv 1-2\nu \simeq \frac{2m^{2}}{H^{2}}< \frac{1}{2}
\end{eqnarray}
 We see that
the departure from scale invariance is still small.
\\\\
\textbf{Critical Mass: $m = \frac{H}{2}$} \\\\
For $m = \frac{H}{2}$, $\nu$ is zero.  Therefore,
\begin{eqnarray}
\lambda_{+} = \lambda_{-} = -\frac{1}{2}
\end{eqnarray}
and $\mathcal{C}_{q}$ and $\mathcal{D}_{q}$ are \textit{$q$-
independent}.  Eq. \eqref{T1} becomes
\begin{eqnarray}\label{T1crit}
\int_{-\infty}^{t_{2}}dt_{1}a_{1} e^{-i
q\tau_{1}}\dot{\mathcal{A}_{p}}(t_{1})
\dot{\mathcal{A}_{p'}}(t_{1}) =-\frac{H}{4}\Big(\mathcal{C}_{p}+
\mathcal{D}_{p}\Big) \Big(\mathcal{C}_{p'}+
\mathcal{D}_{p'}\Big)Ei(-iq\tau_{2})
\end{eqnarray}
Substituting equation above into eq. \eqref{Ttotal}, we therefore
have
\begin{eqnarray}\label{Ttotalcrit}
\mathcal{T} &=& \frac{\pi^{2}H^{2}}{64}|\mathcal{C}_{p}+
\mathcal{D}_{p}|^{2} |\mathcal{C}_{p'}+ \mathcal{D}_{p'}|^{2}
\end{eqnarray}
We see from eqs. \eqref{masscoeff} and \eqref{masscoeff1} that the
coefficients are all momentum independent when $\nu = 0$. Since
\begin{eqnarray}
\Big(\mathcal{C}_{q}+\mathcal{D}_{q}\Big)|_{\nu=0, m= \frac{H}{2}}
= \frac{0}{0}
\end{eqnarray}
we need to use l' Hospital's rule.  From eqs. \eqref{masscoeff}
and \eqref{masscoeff1}, we have
\begin{eqnarray}
\lim_{\nu\rightarrow 0}\Big(\mathcal{C}_{q}+\mathcal{D}_{q}\Big) =
-\frac{i\sqrt{\pi}e^{\frac{i\pi}{4}}}{2(2\pi)^{\frac{3}{2}}\sqrt{H}}
\lim_{\nu\rightarrow 0}\Big[\frac{e^{\frac{i\pi
\nu}{2}}(\frac{2H}{q})^{\nu}}{\sin\nu\pi \Gamma(1-\nu)}-
\frac{e^{-\frac{i\pi \nu}{2}}(\frac{2H}{q})^{-\nu}}{\sin\nu\pi
\Gamma(1+\nu)}\Big]
\end{eqnarray}
Note that
\begin{eqnarray}
\lim_{\nu\rightarrow 0}\frac{e^{\frac{i\pi
\nu}{2}}(\frac{2H}{q})^{\nu}}{\sin\nu\pi \Gamma(1-\nu)} &=&
\lim_{\nu\rightarrow
0}\frac{e^{\frac{i\pi\nu}{2}}(\frac{i\pi}{2}(\frac{2H}{q})^{\nu}+\nu(\frac{2H}{q})^{\nu-1})}
{\pi \cos\nu\pi \Gamma(1-\nu)-\sin\nu \pi
\psi(1-\nu)\Gamma(1-\nu)}\nonumber \\
&=& \frac{i}{2}
\end{eqnarray}
Similarly,
\begin{eqnarray}
\lim_{\nu\rightarrow 0}\frac{e^{-\frac{i\pi
\nu}{2}}(\frac{2H}{q})^{-\nu}}{\sin\nu\pi \Gamma(1+\nu)} &=&
-\frac{i}{2}
\end{eqnarray}
Therefore,
\begin{eqnarray}
\lim_{\nu\rightarrow 0}\Big(\mathcal{C}_{q}+\mathcal{D}_{q}\Big) =
\frac{\sqrt{\pi}e^{\frac{i\pi}{4}}}{2(2\pi)^{\frac{3}{2}}\sqrt{H}}
\end{eqnarray}
We see that the coefficients are $q$-independent.  Hence, eq.
\eqref{Ttotalcrit} becomes
\begin{eqnarray}\label{TTTt}
\mathcal{T} = \frac{\pi^{4}}{1024(2\pi)^{6}}
\end{eqnarray}
Therefore, the correlation function in eq.\eqref{gaugeresultmass1}
becomes
\begin{eqnarray}\label{gaugeresultcritmass}
\int d^{3}x e^{i \textbf{q}\cdot
(\textbf{x}-\textbf{x}')}\Big\langle
\zeta(\textbf{x},t)\zeta(\textbf{x}',t)\Big\rangle &=&
-\frac{\pi^{2}(8\pi G H^{2})^{2}}{1024 q^{3}}
\end{eqnarray}\\
\textbf{Large Mass: $m > \frac{H}{2}$ }
\\\\
For $m > \frac{H}{2}$, $\lambda_{\pm}$ are complex conjugates of
each other as
\begin{eqnarray}
\lambda_{+}\equiv \lambda = -\frac{1}{2}+ i s,  \lambda_{-} =
\lambda^{*}= -\frac{1}{2}- i s
\end{eqnarray}
where $\nu = i s$ and $s \equiv
\sqrt{|\frac{m^{2}}{H^{2}}-\frac{1}{4}|}$ is real.  Notice that
$|\mathcal{C}_{q}|^{2}$ and $|\mathcal{D}_{q}|^{2}$ are\textit{
$q$-independent}.  Therefore, eq. \eqref{T1} becomes
\begin{eqnarray}\label{T1largemass}
\int_{-\infty}^{t_{2}}dt_{1}a_{1} e^{-i
q\tau_{1}}\dot{\mathcal{A}_{p}}(t_{1})
\dot{\mathcal{A}_{p'}}(t_{1})&\rightarrow& \frac{H}{2 i s}\Big[
\lambda^{2} c_{+}a_{2}^{2i s} - \lambda^{*2}c_{-}a_{2}^{-2 i s}
\nonumber \\&&-2 i s |\lambda|^{2}c_{0}Ei(-iq\tau_{2})\Big]
\end{eqnarray}
We see that the time components of the first two terms are complex
conjugates of each other with different constant coefficients.
Integrating over time $t_{2}$ gives
\begin{eqnarray}\label{Tintvec}
\mathcal{T} &=& -\frac{H^{3}}{s }R e\int_{-\infty}^{t}dt_{2}\sin q
\tau_{2}\Big[\lambda^{*2}c^{*}_{+}a_{2}^{-2is}+
\lambda^{2}c^{*}_{-}a_{2}^{2is}+ |\lambda|^{2}c^{*}_{0}\Big]
\nonumber \\&&\times \Big[\lambda^{2}c_{+}a_{2}^{2 i s }-
 \lambda^{*2}c_{-}a_{2}^{-2 i s }-2 i s|\lambda|^{2} c_{0}E i(-i q \tau_{2})\Big]
\nonumber \\&=&\frac{|\lambda|^{4}H^{2}}{s
}\int_{-\infty}^{0}\frac{d \tau_{2}}{\tau_{2}}
 \sin q \tau_{2}\Big(|c_{+}|^{2}-|c_{-}|^{2}- 2 s |c_{0}|^{2}S i q \tau_{2} \Big)
\end{eqnarray}
where the other terms vanish in the limit of $t\rightarrow \infty$
because of the oscillating behavior of the integrand $e^{\pm i H
t_{2}}$.  We can integrate further with the use of Mathematica
\begin{eqnarray}
\int_{-\infty}^{0}\frac{d\tau_{2}}{\tau_{2}}\sin q\tau_{2}Si
q\tau_{2} = -\frac{\pi^{2}}{8}\\
\int_{-\infty}^{0}\frac{d\tau_{2}}{\tau_{2}}\sin q\tau_{2} =
\frac{\pi}{2}
\end{eqnarray}
Therefore, eq. \eqref{Tintvec} becomes
\begin{eqnarray}\label{Tintvec1}
\mathcal{T} &=&\frac{\pi|\lambda|^{4}H^{2}}{4s }
\Big[2(|c_{+}|^{2}-|c_{-}|^{2})
 + \pi s|c_{0}|^{2}\Big]
\end{eqnarray}
We now need to calculate what $|c_{\pm,0}|^{2}$ are.  From eq.
\eqref{c0pm}, we have
\begin{eqnarray}
|c_{+}|^{2} &=& |\mathcal{C}_{p}|^{2}|\mathcal{C}_{p'}|^{2}\\
|c_{-}|^{2} &=& |\mathcal{D}_{p}|^{2}|\mathcal{D}_{p'}|^{2}\\
|c_{0}|^{2} &=& |\mathcal{C}_{p}\mathcal{D}_{p'}+
\mathcal{D}_{p}\mathcal{C}_{p'}|^{2}
\end{eqnarray}
From eqs. \eqref{masscoeff} and \eqref{masscoeff1}, we have
\begin{equation}\label{masscoeffsq}
|\mathcal{C}_{q}|^{2} = \frac{\pi}{4(2\pi)^{3} H|\sin \pi\nu
|^{2}}\frac{e^{-\pi s}}{|\Gamma(1-\nu)|^{2}}
\end{equation}
and
\begin{equation}\label{masscoeff1sq}
|\mathcal{D}_{q}|^{2} =\frac{\pi}{4(2\pi)^{3}H|\sin\nu\pi|^{2}}
\frac{e^{\pi s}}{|\Gamma(1+\nu)|^{2}}
\end{equation}
 We can see that $|\mathcal{C}_{q}|^{2}$ and $|\mathcal{D}_{q}|^{2}$
  are \textit{momentum independent} because
 $\nu \equiv i s$ is purely imaginary in the large mass limit when $m
> \frac{ H}{2}$.  Therefore,
\begin{eqnarray}
|c_{+}|^{2}-|c_{-}|^{2} = \frac{\pi^{2}(e^{-2\pi s}- e^{2\pi
s})}{16(2\pi)^{6}H^{2}|\sin\nu\pi|^{4}|\Gamma(1+\nu)|^{4}}
\end{eqnarray}
Using $|\Gamma(1+i s)|^{2}= |\Gamma(1-i s )|^{2}= \frac{\pi s}{
\sinh \pi s}$ and $\sin i s = i\sinh s$ for real $s$, the equation
above is simplified as
\begin{eqnarray}\label{cpmvec}
|c_{+}|^{2}-|c_{-}|^{2} = -\frac{\coth \pi s
}{4(2\pi)^{6}s^{2}H^{2}}
\end{eqnarray}
where we use $\sinh 2 x = 2\sinh x \cosh x$.  Also,
\begin{eqnarray}\label{c0vec}
|c_{0}|^{2} = \frac{ \cos^{2} (s \ln \frac{p}{p'})
}{4(2\pi)^{6}s^{2}H^{2}\sinh^{2}\pi s}
\end{eqnarray}
Notice that the coefficients $|c_{+}|^{2}-|c_{-}|^{2}$ are
completely momentum independent and $|c_{0}|^{2}$ is nearly
momentum independent ($\ln
\frac{|\textbf{p}|}{|\textbf{p}+\textbf{q}|}\rightarrow 0$ when
$q\rightarrow 0$).  Substituting eqs. \eqref{cpmvec}, and
\eqref{c0vec} into eq. \eqref{Tintvec1}, we have
\begin{eqnarray}\label{Tintvec2}
\mathcal{T} &=&\frac{\pi|\lambda|^{4}}{8s(2\pi)^{6} } \Big[
-\frac{\coth \pi s }{s^{2}}
 +  \frac{\pi\cos^{2} (s \ln \frac{p}{p'})
}{2s\sinh^{2}\pi s}\Big]
\end{eqnarray}
From eqs. \eqref{gaugeresultmass1} and  \eqref{Tintvec2}, we have
\begin{eqnarray}\label{resultvec}
\int d^{3}x e^{i \textbf{q}\cdot
(\textbf{x}-\textbf{x'})}\Big\langle
\zeta(\textbf{x},t)\zeta(\textbf{x}',t)\Big\rangle_{loop} =
-\frac{3\pi(8\pi G H^{2})^{2}|\lambda|^{4}}{32
s(2\pi)^{3}q^{6}}\int d^{3}pd^{3}p' \nonumber \\\times
\delta^{3}(\textbf{q}+\textbf{p}+\textbf{p}') \Big[ -\frac{\coth
\pi s }{s^{2}}
 +  \frac{\pi\cos^{2} (s\ln \frac{p}{p'})}{2s\sinh^{2}\pi s}\Big]
\end{eqnarray}
The result of the momentum integrals $p,p'$ gives the momentum
dependence as $q^{3}$, which cancels with the $q^{-6}$ factor in
eq. \eqref{resultvec}.  We therefore have the approximated scale
invariant spectrum after horizon exit as
\begin{eqnarray}\label{resultvec1}
\int d^{3}x e^{i \textbf{q}\cdot
(\textbf{x}-\textbf{x'})}\Big\langle
\zeta(\textbf{x},t)\zeta(\textbf{x}',t)\Big\rangle_{loop, m>
\frac{H}{2}}= \frac{3\pi^{2}(8\pi G H^{2})^{2}|\lambda|^{4}}{8
s^{3}(2\pi)^{3}q^{3}} \coth \pi s
\end{eqnarray}
where $\lambda = -\frac{1}{2}+ i s$ and $s =
\sqrt{|\frac{m^{2}}{H^{2}}-\frac{1}{4}|}$.
\vspace{12pt}
\begin{center}
{\bf VIII.  CONFORMAL SCALAR LOOP, INFLATON, AND GRAVITY}
\end{center}
\nopagebreak We have learned that the spectrums of massless
minimal coupled scalar, massless fermion, and massless vector
fields loops \textit{all} go as $(8\pi G H^{2})^{2}q^{-3}\ln q$.
We would like to investigate whether this is also true for
conformal scalar loop.  The full action considered during
inflation is
 \begin{eqnarray}\label{conaction}
\mathcal{L}  = -\frac{1}{2}\sqrt{-g}\Big[\frac{1}{8\pi G}R
+g^{\mu\nu}\partial_{\mu}\varphi\partial_{\nu}\varphi +
2V(\varphi) + g^{\mu\nu}\partial_{\mu}\chi\partial_{\nu}\chi - \xi
R\chi^{2}\Big]
\end{eqnarray}
where $\varphi$ is an inflaton , $\chi$ is additional conformal
scalar matter in which $\langle\chi\rangle = 0$, and $\xi =
\frac{1}{6},  0$ for conformal and minimal couplings respectively.
We consider this to see how the conformal scalar affects the
spectrum $\langle\zeta\zeta\rangle$ through the interaction with
gravitational fluctuation.  To arrive at the field equation of the
conformal scalar field, we
 need the action up to the second order in the field fluctuations
 which is
 \begin{eqnarray}
\mathcal{L}_{\chi}^{(2)} =
 \frac{a^{3}}{2}\Big(\dot{\chi}^{2}- \frac{(\partial_{i}\chi)^{2}}{a^{2}}
 -12\xi H^{2}\chi^{2}\Big)
 \end{eqnarray}
 where $\bar{R} = -12 H^{2}$.
 Varying the second order of the action with respect to $\chi$, we
 have the field equation of the conformal scalar field as
 \begin{eqnarray}\label{confe}
 \ddot{\chi}_{q}+ 3 H\dot{\chi}_{q} + \Big(\frac{q^{2}}{a^{2}}+
 12\xi
 H^{2}\Big)\chi_{q} = 0
 \end{eqnarray}
 Notice that if there is no extra term $12\xi H^{2}\chi$ (or $\xi = 0$), this is just
 a minimal coupled massless scalar, in which the dominant solution
 approaches a constant at late time.  It is known that the massless minimal
 coupled scalar produces a scale free spectrum.  We would like to
 investigate the momentum dependence of the power spectrum due to
 conformal scalar loops here when $\xi = \frac{1}{6}$.
 Eq.  \eqref{confe} can be solved exactly by re-scaling the field
 $\chi \equiv u/a$.  Hence, for the conformal scalar,
 \begin{eqnarray}\label{confe1}
 u_{q}'' +  \Big(q^{2}- \frac{a''}{a} + 2 H^{2} a^{2}
 \Big)u_{q} = 0
 \end{eqnarray}
During inflation, $a \simeq -\frac{1}{H\tau}$, therefore, the last
two terms of eq. \eqref{confe1} are cancelled.  We arrive at a
simple field equation of the conformal scalar field
\begin{eqnarray}\label{confe2}
 u_{q}'' +  q^{2}u_{q} = 0
 \end{eqnarray}
 Therefore, the solution to the equation above is just a simple plane
 wave valid to all wavelengths
 \begin{eqnarray}\label{sol}
 \chi_{q}(t) = u_{q}(t)/a(t) = \frac{1}{(2\pi)^{\frac{3}{2}}a(t)\sqrt{2 q}}e^{-i q\tau}
 \end{eqnarray}
 where we choose the constant coefficients to match with the
 positive mode solution inside the horizon.  From eq.  \eqref{sol},
 the conformal scalar field correlation function
 to leading order is
 \begin{eqnarray}
 \Big\langle \chi(\textbf{x},t)\chi(\textbf{x}',t)\Big\rangle
 &=& \int d^{3}q e^{i
 \textbf{q}\cdot(\textbf{x}-\textbf{x}')}|\chi_{q}(t)|^{2} \nonumber \\
 &=& \int \frac{d^{3}q}{(2\pi)^{3}} e^{i
 \textbf{q}\cdot(\textbf{x}-\textbf{x}')}\frac{1}{2 q a^{2}(t)}
 \end{eqnarray}
We see that its momentum dependence is far from scale invariant at
the classical level.  However, we never observe the product of the
scalar fields in CMB anisotropy but rather the correlation
function of the temperature or density fluctuations which is
related  to the conserved quantity $\zeta$.  Therefore, we study
how the conformal scalar field affects the observable power
spectrum $\langle\zeta\zeta\rangle$ via the gravitational
interactions at the quantum level.  We can calculate the trilinear
vertices due to the conformal scalar $\chi$ and gravity $\zeta$.
They are
 \begin{eqnarray}\label{conactionexpand}
 \mathcal{L}_{\zeta\chi\chi} &=&
 -\frac{a}{2}\zeta(\partial_{i}\chi)^{2}
 + a \partial_{i}\Big(\frac{\zeta}{H}- \epsilon H
 a^{2}\nabla^{-2}\dot{\zeta}\Big)\dot{\chi}\partial_{i}\chi \nonumber \\&&
 -\frac{a}{2 H}\dot{\zeta}(\partial_{i}\chi)^{2}-
 \frac{a^{3}}{2H}\dot{\zeta}\dot{\chi}^{2}+ \frac{3
 a^{3}}{2}\zeta\dot{\chi}^{2}\nonumber \\&&
 - a^{3}\Big(3H^{2}\zeta + H\dot{\zeta}-\frac{\delta R}{12}\Big)\chi^{2}
 \end{eqnarray}
 where the last line above contains the additional terms from the
 massless minimal coupled
 scalar.  Those terms in the last line arise from the conformal term which is
 $\frac{\sqrt{-g}}{12}R\chi^{2}$.  We see that the
 interactions above are rather complicated.
As derived in the appendix, the interaction vertices
 can be written in a more compact form  as
\begin{eqnarray}
H_{\zeta\chi\chi}(t) = -\int d^{3}x \epsilon H a^{5} (T^{00}+
a^{2}T^{ii})\nabla^{-2}\dot{\zeta}
\end{eqnarray}
where $T^{\mu\nu}$ is the energy momentum tensor at the second
order of arbitrary matter.
 The combination of time and space components of the energy momentum
 tensor is [14]
\begin{eqnarray}
T^{00}+ a^{2}T^{ii} &=& 2(1-3\xi) \dot{\chi}^{2}+
12\xi^{2}H^{2}\chi^{2} +
\frac{2\xi}{a^{2}}\chi_{;i}^{2}-2\xi\chi\Big(
\chi_{;00}+\frac{\chi_{;ii}}{a^{2}} \Big) \nonumber \\
&&- \xi \chi^{2}\Big( \bar{R}_{00}+\frac{\bar{R}_{ii}}{a^{2}}-
\bar{R} + 3\xi \bar{R}\Big)
\end{eqnarray}
We can check that $T^{00}+a^{2}T^{ii} = 2\dot{\chi}^{2}$ for
minimal coupling $\xi = 0$.  For conformal coupling $\xi =
\frac{1}{6}$, we have
\begin{eqnarray}
T^{00}+ a^{2}T^{ii} = \dot{\chi}^{2}+ \frac{1}{3}\Big(
\frac{(\partial_{i}\chi)^{2}}{a^{2}}-2\chi\ddot{\chi}-H^{2}\chi^{2}\Big)
\end{eqnarray}
where the conformal field equation \eqref{confe} is used and
\begin{eqnarray}
\bar{R}_{00}+\frac{\bar{R}_{ii}}{a^{2}}- \bar{R} + 3\xi \bar{R} =
0
\end{eqnarray}
due to $\bar{R}_{00} = 3H^{2}, \bar{R}_{ii} = -9a^{2}H^{2}$ and
$\bar{R}= -12H^{2}$ in de-Sitter phase inflation.  Therefore, the
trilinear interaction Hamiltonian of the conformal scalar $\chi$
and gravity $\zeta$ is
\begin{eqnarray}\label{conactionexpandsim}
H_{\zeta \chi\chi}(t) = -\int d^{3}x \epsilon H a^{5}
\Big[\dot{\chi}^{2}+ \frac{1}{3}(
\frac{(\partial_{i}\chi)^{2}}{a^{2}}-2\chi\ddot{\chi}
-H^{2}\chi^{2})\Big]\nabla^{-2}\dot{\zeta}
\end{eqnarray}
To calculate the loop spectrum in the commutator
$[H_{1},[H_{2},Q]]$, we can use the general formula in
eq.\eqref{genformula} with the replacement
\begin{eqnarray}
\Psi^{*}\Psi &\rightarrow& \dot{\chi}^{2}+ \frac{1}{3}\Big(
\frac{(\partial_{i}\chi)^{2}}{a^{2}}-2\chi\ddot{\chi}-H^{2}\chi^{2}\Big)\\
\zeta_{q}(t_{1,2})&\rightarrow& -\dot{\zeta}_{q}(t_{1,2})/q^{2}\\
V(t) &=& -\epsilon H a^{5}(t)
\end{eqnarray}
Hence,
\begin{eqnarray}\label{Mchi}
\mathcal{M}_{\chi} &\equiv& \pi^{*}(t_{2})\pi(t_{1})
\end{eqnarray}
where
\begin{eqnarray}
\pi(t) =
\dot{\chi}_{p'}\dot{\chi}_{p}-\frac{1}{3}\chi_{p'}\ddot{\chi}_{p}
-\frac{1}{3}\ddot{\chi}_{p'}\chi_{p}-\frac{1}{3}\Big(
\frac{pp'}{a^{2}}+H^{2}\Big)\chi_{p'}\chi_{p}
\end{eqnarray}
With
\begin{eqnarray}
\dot{\chi}_{q}(t) &=& -\Big(H+ \frac{iq}{a}\Big)\chi_{q}\\
 \ddot{\chi}_{q}(t)
&=& \Big[H^{2}+ 3i H\frac{q}{a}-\frac{q^{2}}{a^{2}}\Big]\chi_{q}
\end{eqnarray}
many terms are cancelled.  Therefore,
\begin{eqnarray}\label{pi1}
\pi(t) = \frac{1}{3 a^{2}}\Big[ p^{2}+p'^{2}-4 p
p'\Big]\chi_{p}(t)\chi_{p'}(t)
\end{eqnarray}
Hence,
\begin{eqnarray}\label{Mchina}
\mathcal{M}_{\chi}  &=&
 \frac{1}{9 a_{1}^{2}a_{2}^{2}}\Big( p^{2}+p'^{2}-4 p
p'\Big)^{2}\chi^{*}_{p}(t_{2})\chi^{*}_{p'}(t_{2})\chi_{p}(t_{1})\chi_{p'}(t_{1})
\nonumber \\
&=& \frac{1}{36(2\pi)^{6} a_{1}^{4}a_{2}^{4}p p'}\Big(
p^{2}+p'^{2}-4 p p'\Big)^{2}e^{-i(p+p')(\tau_{1}-\tau_{2})}
\end{eqnarray}
From eq. \eqref{genformula}, we have
\begin{eqnarray}\label{conresult}
\int d^{3}x e^{i \textbf{q}\cdot
(\textbf{x}-\textbf{x}')}\Big\langle
\zeta(\textbf{x},t)\zeta(\textbf{x}',t)\Big\rangle_{loop} =
-8(2\pi)^{9}\int
d^{3}pd^{3}p' \delta^{3}(\textbf{q}+\textbf{p}+\textbf{p}')\nonumber \\
\times\int_{-\infty}^{t}dt_{2}\epsilon_{2}H_{2}a_{2}^{5}
\int_{-\infty}^{t_{2}}dt_{1}\epsilon_{1}H_{1}a_{1}^{5}
Re\Big(\mathcal{Z}\pi^{*}_{2}\pi_{1}\Big)
\end{eqnarray}
where $\mathcal{\zeta}$ remains the same as in eq.
\eqref{zetapart}.  Substituting the $\mathcal{Z}$ part eq.
\eqref{zetapart} and matter part eq.  \eqref{pi1} into eq.
\eqref{conresult}, we have the correlation function due to the
conformal scalar field loop
\begin{eqnarray}\label{conresultx}
&&\int d^{3}x e^{i \textbf{q}\cdot
(\textbf{x}-\textbf{x}')}\Big\langle
\zeta(\textbf{x},t)\zeta(\textbf{x}',t)\Big\rangle_{loop} =
-\frac{2(2\pi)^{3}(8\pi G H^{2})^{2}}{9 q^{6}}\int d^{3}pd^{3}p'
\nonumber \\&&\times \delta^{3}(\textbf{q}+\textbf{p}+\textbf{p}')
\Big[ p^{2}+p'^{2}-4 p p'\Big]^{2}
Re\int_{-\infty}^{\infty}dt_{2}a_{2}(e^{-iq\tau_{2}}-e^{iq\tau_{2}})
\chi^{*}_{p}(t_{2})\chi^{*}_{p'}(t_{2})\nonumber \\&&\times
\int_{-\infty}^{t_{2}}dt_{1}a_{1}e^{-iq\tau_{1}}\chi_{p}(t_{1})\chi_{p}(t_{1})
\end{eqnarray}
With the mode solution in eq. \eqref{sol} and the result of the
time integrations, we have the loop correlation function as
\begin{eqnarray}\label{conresult2}
\int d^{3}x e^{i \textbf{q}\cdot
(\textbf{x}-\textbf{x'})}\Big\langle
\zeta(\textbf{x},t)\zeta(\textbf{x}',t)\Big\rangle_{loop} =
\frac{(8\pi G H^{2})^{2}}{36(2\pi)^{3}q^{7}}\nonumber \\\times\int
d^{3}pd^{3}p' \delta^{3}(\textbf{q}+\textbf{p}+\textbf{p}') \frac{
\Big( p^{2}+p'^{2}-4 p p'\Big)^{2}}{pp'(p+p'+q)}
\end{eqnarray}
With the dimensional regularization as done before, we have the
conformal scalar loop correlation function as
\begin{eqnarray}\label{condimpresult}
\int d^{3}x e^{i \textbf{q}\cdot
(\textbf{x}-\textbf{x}')}\Big\langle
\zeta(\textbf{x},t)\zeta(\textbf{x}',t)\Big\rangle_{loop} =
-\frac{\pi(8\pi G H^{2})^{2}}{90(2\pi)^{3}q^{3}}\Big[\ln q +
C\Big]
\end{eqnarray}
We see that it is nearly scale invariance and smaller than the
classical result by a factor of $\frac{|\epsilon|H^{2}\ln
q}{M_{Pl}^{2}}$.
\vspace{12pt}
\begin{center}
{\bf IX.  SUMMARY OF ALL RESULTS}
\end{center}
\nopagebreak We study quantum effects of cosmological correlations
due to the interactions of gravitational and matter fluctuations.
It is shown that departures from scale invariance are never large,
regardless of what kind of theories, what kind of matter, or what
kind of inflaton potential $V(\varphi)$ are used. \\
The results in this paper may be compared with the Weinberg's
result [2],
\\
\textbf{Minimal Coupled Scalar Field Loops}
\begin{eqnarray}\label{Smlcorr}
\langle \zeta\zeta\rangle_{m =0} = - \frac{ \pi (8\pi G)^{2}
H(t_{q})^{4}}{ 15(2\pi)^{3} q^{3}}\Big[\ln q + C\Big]
\end{eqnarray}
where $H(t_{q})\propto q^{-\epsilon}$ is the expansion rate at the
time of horizon exit.\\\\
The results in this paper are
\\
\\
\textbf{Dirac Field Loops}
\begin{eqnarray}\label{Smlcorr}
\langle \zeta\zeta\rangle_{m_{f}=0} =- \frac{ 4\pi (8\pi G)^{2}
H(t_{q})^{4}}{ 15(2\pi)^{3} q^{3}}\Big[\ln q + C\Big]
\end{eqnarray}
\begin{eqnarray}\label{SZetaInIn}
\langle \zeta\zeta\rangle_{m_{f}\neq 0} \rightarrow -\frac{7(8\pi
G)^{2}m_{f}^{2}H(t_{q})^{2}}{24 q^{3}\cosh^{2}\frac{\pi m_{f}}{H}}
\end{eqnarray}
\textbf{Gauge Field Loops}
\begin{eqnarray}\label{Sgaugedimpresult}
\langle \zeta\zeta\rangle_{m_{v}=0} = -\frac{ 44\pi(8\pi G)^{2}
H(t_{q})^{4}}{30(2\pi)^{3}q^{3}} \Big[\ln q + C\Big]
\end{eqnarray}
\begin{eqnarray}\label{vecresultSM5}
\langle \zeta\zeta\rangle_{m_{v}< \frac{H}{2}}\rightarrow-\frac{24
\Gamma^{2}(\nu)\lambda_{-} \lambda_{+}^{3}(8\pi G
H(t_{q})^{2})^{2}}{(2\pi)^{3}[2 H(t_{q})
a(t)]^{\eta}}\frac{(1-\eta\ln q\tau)}{(1-\eta)(2+\eta)\eta^{2}
q^{3-\eta}}
\end{eqnarray}
 where $\lambda_{\pm}= -\frac{1}{2}\pm \nu$,
  $0 \ll 2\nu = \sqrt{1 - \frac{4 m_{v}^{2}}{H^{2}}}< 1$ and $\eta =
 1-2\nu < 0.5$.
\begin{eqnarray}\label{Smovec}
\langle\zeta\zeta\rangle_{m_{v} = \frac{H}{2}}
&=&-\frac{\pi^{2}(8\pi G)^{2}H(t_{q})^{4}}{1024 q^{3}}
\end{eqnarray}
\begin{eqnarray}\label{Sresultvec1}
\langle \zeta\zeta\rangle_{m_{v} > \frac{H}{2}}\rightarrow
\frac{3\pi^{2}(8\pi G)^{2}H(t_{q})^{4}|\lambda|^{4}}{8
s^{3}(2\pi)^{3}q^{3}} \coth \pi s
\end{eqnarray}
where $\lambda = -\frac{1}{2}+ i s$ and $s
=\sqrt{|\frac{m_{v}^{2}}{H^{2}}-\frac{1}{4}|}$. \\
\begin{eqnarray}\label{Sresultvec1Mpl}
\langle \zeta\zeta\rangle_{m_{v}=M_{Pl}}\rightarrow
\frac{3\pi^{2}(8\pi G)^{2}M_{Pl}H(t_{q})^{3}}{8(2\pi)^{3}q^{3}}
\end{eqnarray}
\textbf{Conformal Scalar Field Loops}
\begin{eqnarray}\label{Scondimpresult}
\langle \zeta\zeta\rangle = -\frac{\pi(8\pi G)^{2}
H(t_{q})^{4}}{90(2\pi)^{3}q^{3}}\Big[\ln q + C\Big]
\end{eqnarray}
We see that even when the mass is as large as $M_{Pl}$, the
one-loop result is still naturally smaller than the classical one.
Therefore, no fine tuning is needed.  The result above is still
valid in the realistic and the general potential $V(\varphi,
\bar{\psi}\psi, A_{\mu}A^{\mu})$.  The reason is that we choose
the gauge in which the inflaton does not
fluctuate [7] ($\delta\varphi = 0$). \\

Even when the additional interactions of inflaton and matters
arise, the results in eqs. \eqref{SZetaInIn} and
\eqref{Sresultvec1} do not change but rather the masses are
shifted by
\begin{eqnarray}
V(\varphi)&\rightarrow& V(\varphi, \bar{\psi}\psi,
A_{\mu}A^{\mu})\\
 m_{f} &\rightarrow& m_{f} +
\frac{\partial^{2}V}{\partial
\bar{\psi}\partial\psi}|_{\psi = 0}\\
m^{2}_{v} &\rightarrow& m_{v}^{2} + \frac{\partial^{2}V}{\partial
A_{\mu} \partial A^{\mu}}|_{A_{\mu} = 0}
\end{eqnarray}
The is because the mass shift( $\frac{\partial^{2}V}{\partial
\bar{\psi} \partial\psi}|_{\psi =0}, \frac{\partial^{2}V}{\partial
A_{\mu}\partial A^{\mu}}|_{A_{\mu}=0}$), which is a function of
the unperturbed inflaton only, does not change much during
inflation. We therefore can approximate the unperturbed inflaton
at the time of horizon exit $\bar{\varphi}(t)\simeq
\bar{\varphi}(t_{q})$. Hence there is no additional consequence to
the momentum dependence of loop
spectrums.\\

Therefore, the spectrums are nearly scale invariant even if we add
interactions of arbitrary matter and the inflaton to the
interactions of matter and gravity.  These results imply that we
and the things around us did not come from nothing or an unknown
scalar field as in conventional beliefs.  Rather it points to the
fact that we originated from quantum fluctuations due to the
interactions between gravity and various matters during the time
of Big Bang inflation.

\vspace{12pt}
\begin{center}
{\bf ACKNOWLEDGMENTS}
\end{center}
For helpful conversations I am grateful to my supervisor S.
Weinberg.  I also thank the referee for remarks that to a clearer
presentation and for pointing out some typos.  I thank A. Fassi,
C. Hong, and S. Young for correcting my English. This material is
based upon work supported by the National Science Foundation under
Grant No. PHY-$0455649$.

\vspace{12pt}
\begin{center}
{\bf APPENDIX: HIGHER ORDER FLUCTUATIONS}
\end{center}
\nopagebreak
This appendix is to clarify and derive interactions
of matter and gravitational fluctuations used in the loop
calculations.  The method of expansion and quantization shown by
Weinberg [2] has a more compact form than the direct expansion of
matter and gravitational fluctuations.  Following his method, we
can extend the calculation to other matter such as fermion, gauge,
and conformal scalar fields without many difficulties.  We would
like to show the calculation in detail for the general reader.
\\

In cosmological fluctuations, we generally expand the gravity and
an inflaton around a time dependent background such that
\begin{eqnarray}
g_{\mu\nu}(\textbf{x},t) = \bar{g}_{\mu\nu}(t)+\delta g_{\mu\nu}(\textbf{x},t)\\
\varphi(\textbf{x},t) =
\bar{\varphi}(t)+\delta\varphi(\textbf{x},t)
\end{eqnarray}
When we add any other kinds of matter which have unbroken
symmetries, they can be expanded as
\begin{eqnarray}
M(\textbf{x},t) = 0 + \delta M(\textbf{x},t)
\end{eqnarray}
where $M$ represents any additional matter such as fermions,
vector bosons, and conformal scalar fields.  The perturbation to
the metric around an FRW background can always be placed in the
form of
\begin{equation}\label{A2}
ds^{2} = -(1+E)dt^{2} + 2 a(t) F_{,i} dt dx^{i} +
a^{2}(t)((1+A)\delta_{ij}+ B_{,ij})dx^{i}dx^{j}
\end{equation}
where we only consider the scalar mode which is the subject of
interest here.  The gauge invariant observable quantity is defined
as
\begin{eqnarray}
\zeta_{q} \equiv \frac{A_{q}}{2}-\frac{H\delta
\varphi_{q}}{\dot{\bar{\varphi}}}
\end{eqnarray}
to linear order.  We see that $\zeta$ is a gauge invariant
quantity that relates to both matter and gravitational
fluctuations.  There is a need for us to
learn how to quantize such theories with minimum complication.\\

Since the inflaton and gravity are related through Einstein's
equation, we have some choices in choosing a gauge.  It is found
to be more convenient to choose a gauge such that the inflaton
does not fluctuate [7].  Therefore, we can write down all the
components of the gravitational fluctuations $\delta g_{\mu\nu}$
in terms of a single variable $\zeta$ by solving Einstein's
equation in the Maldacena gauge $\delta\varphi = B =  0$.  From
the gravitational field equations and the energy conservation
equations[9],
\begin{eqnarray}
0 &=& \dot{A}-HE \\
0 &=& H \dot{E} + 2(3H^{2}+ \dot{H})E+ a^{-2}\nabla^{2}A-
\ddot{A}-6H\dot{A}+ 2a^{-1}H\nabla^{2}F \\
0 &=& -\frac{1}{2}\frac{d}{dt}(E\dot{H})-
3H\dot{H}E-a^{-1}\dot{H}\nabla^{2}F + \frac{3}{2}\dot{H}\dot{A}
\end{eqnarray}
Solving the equations above, we therefore have
\begin{eqnarray}
A = 2\zeta,  E = \frac{2\dot{\zeta}}{H},  F = -\frac{\zeta}{a H}+
\epsilon a \nabla^{-2}\dot{\zeta}
\end{eqnarray}
where $\epsilon \equiv -\frac{\dot{H}}{H^{2}}$.  By eliminating
$E$ and $F$ yields a differential equation for $A$:
\begin{eqnarray}
\ddot{A} +\Big(3H - \frac{2\dot{H}}{H}+
\frac{\ddot{H}}{\dot{H}}\Big)\dot{A}-\frac{\nabla^{2}A}{a^{2}}= 0
\end{eqnarray}
This is sometimes known as Mukhanov equation[1], in which $A =
2\zeta$.  We can write the metric and its fluctuations in terms of
$\zeta$ as
\begin{eqnarray}
g_{00} &=& -\Big(1+\frac{2\dot{\zeta}}{H}\Big)=  N_{i}N^{i}-N^{2}\\
g_{0i} &=& \partial_{i}\Big(-\frac{\zeta}{H}+ \epsilon a^{2}\nabla^{-2}\dot{\zeta}\Big)= N_{i} \\
g_{ij} &=& a^{2}\delta_{ij}\Big(1+ 2\zeta\Big) = h_{ij}
\end{eqnarray}
The determinant of the metric is
\begin{equation}
\sqrt{-g} = N\sqrt{h} = a^{3}(1+\frac{\dot{\zeta}}{H})e^{3\zeta}
\end{equation}
The gravitational, inflaton, and matter actions in $\delta\varphi
=0$ gauge are
\begin{eqnarray}
\mathcal{L} = \frac{\sqrt{-g}}{2}\Big[ \dot{\bar{\varphi}}^{2}+
2V(\bar{\varphi})+ \frac{R}{8\pi G} \Big] + \mathcal{L}_{M}(\chi,
\bar{\psi}\psi, A_{\mu}A^{\mu})
\end{eqnarray}
where $\mathcal{L}_{M}(\chi, \bar{\psi}\psi, A_{\mu}A^{\mu})$ are
the additional types of matter such as the conformal scalar,
fermion, and vector bosons that do not have the background.  The
first three terms give vertices of purely $\zeta$.  We are
presently interested in the interactions of matter and
gravitational fluctuations in the last term ($\mathcal{L}_{M}$
term) because, in general, the matter loops are larger than the
$\zeta$ loops by a factor of $8\pi G$.  Therefore, the time
dependent tri-linear vertices of general matter are
\begin{eqnarray}\label{htriint}
H_{\zeta M M}(t) &=& -\frac{1}{2}\int d^{3}x a^{3}T^{\mu\nu}\delta
g_{\mu\nu}\nonumber \\
&=& \int d^{3}x a^{3}\Big(
\frac{\dot{\zeta}}{H}T^{00}-\partial_{i}(-\frac{\zeta}{H}+\epsilon
a^{2}\nabla^{-2}\dot{\zeta})T^{0i}-a^{2}\zeta T^{ii}\Big)
\end{eqnarray}
where $T^{\mu\nu}$ is the energy momentum tensor of arbitrary
matter evaluated at quadratic order in fluctuations.  With the
Bianchi Identity,
\begin{eqnarray}
T^{\mu\nu}_{;\nu} = T^{\mu\nu}_{,\mu}+
\Gamma^{\mu}_{\mu\lambda}T^{\lambda\nu}+
\Gamma^{\nu}_{\mu\lambda}T^{\mu\lambda}=0
\end{eqnarray}
we have
\begin{eqnarray}\label{bian}
\frac{1}{a^{3}}\frac{d}{dt}(a^{3}T^{00})+
a\dot{a}T^{ii}+\partial_{i}T^{i0}= 0
\end{eqnarray}
where we use $\bar{\Gamma}^{i}_{i0} = 3 H, \bar{\Gamma}^{0}_{ij} =
a\dot{a}\delta_{ij},
\bar{\Gamma}^{0}_{0\lambda}=\bar{\Gamma}^{0}_{i0}=\bar{\Gamma}^{i}_{ij}=0$
for the unperturbed FRW metric.  Integrating by parts in space and
using the Bianchi Identity eq.  \eqref{bian}, eq. \eqref{htriint}
becomes
\begin{eqnarray}\label{htriint1}
H_{\zeta M M}(t) &=& \int d^{3}x a^{3}\Big(
\frac{\dot{\zeta}}{H}T^{00}+ (\frac{\zeta}{H
a^{3}}-\frac{\epsilon}{a}\nabla^{-2}\dot{\zeta})\frac{d}{dt}(a^{3}T^{00})
\nonumber
\\&&-\epsilon
H a^{2}\nabla^{-2}\dot{\zeta}( a^{2}T^{ii})\Big)
\end{eqnarray}
where the term $a^{2}\zeta T^{ii}$ is cancelled.  With the
Mukhanov equation
\begin{eqnarray}
\ddot{\zeta}+ [3H+
\frac{\dot{\epsilon}}{\epsilon}]\dot{\zeta}-\frac{\nabla^{2}}{a^{2}}\zeta
= 0
\end{eqnarray}
Eq. \eqref{htriint1} is simplified as
\begin{equation}\label{htriint2}
H_{\zeta M M}(t) = Z(t) + \dot{Y}(t)
\end{equation}
where
\begin{eqnarray}\label{ZY}
Z(t) &=& -\int d^{3}x \epsilon H a^{5}
(T^{00}+a^{2}T^{ii})\nabla^{-2}\dot{\zeta} \\
 Y(t) &=&
a^{6}T^{00}\Big(\frac{\zeta}{Ha^{3}}-\frac{\epsilon}{a}\nabla^{-2}\dot{\zeta}\Big)
\end{eqnarray}
The term $\dot{Y}(t)$ can be removed by the field redefinition of
$\tilde{\zeta} \equiv \exp(-iY)\zeta\exp(i Y)$ as mentioned in
[2].  To see more clearly, for any interaction Hamiltonian of the
the form \eqref{htriint2}, Eq. \eqref{SWInInfor} can be put in the
form
\begin{eqnarray}
\langle Q(t)\rangle &=& \sum_{N=0}^{\infty}i^{N}
\int_{-\infty}^{t} dt_{N}\int_{-\infty}^{t_{N}}
dt_{N-1}...\int_{-\infty}^{t_{2}} dt_{1}\nonumber \\&&\times
\langle[\tilde{H}_{I}(t_{1}),[\tilde{H}_{I}(t_{2}),...[\tilde{H}_{I}(t_{N}),
\tilde{Q}^{I}(t)]...]]]\rangle
\end{eqnarray}
where
\begin{eqnarray}
\tilde{H}_{I}(t) &=& e^{iY(t)}\Big[ Z(t) +\dot{Y}(t) +i
e^{-iY(t)}\Big(\frac{d}{d t}e^{iY(t)}\Big) \Big]e^{-iY(t)}
\nonumber \\&=& Z(t) +
i[Y(t),Z(t)]+\frac{i}{2}[Y(t),\dot{Y}(t)]+...
\end{eqnarray}
and
\begin{eqnarray}
\tilde{Q}^{I}(t)&=& e^{iY(t)}Q^{I}(t)e^{-iY(t)}\nonumber \\&=&
Q^{I}(t)+ i[Y(t), Q^{I}(t)]- \frac{1}{2}[Y(t),[Y(t),Q^{I}(t)]]+...
\end{eqnarray}
As mentioned in [2], the redefinition of the operators is
necessary.  It is only products of the redefined field operators
whose expectation values may be expected to give results that
converge at late times.  The results contributed from $Y(t)$ part
only give a sum of powers of $q$ with divergent coefficients, but
with no logarithmic singularity in $q$.  Therefore, $\zeta$ used
in the sections IV-VIII is a new redefined variable, in which we
can safely calculate the contribution from the $Z(t)$ part only.
The calculation in this way is more simplified than the direct
expansion of the fluctuations. \vspace{12pt}
\begin{center}
{\bf REFERENCES}
\end{center}
\nopagebreak

\end{document}